\documentclass[%
reprint,
 amsmath,amssymb,
aps,
prx,
]{revtex4-2}

\usepackage{dcolumn}
\usepackage{bm}
\usepackage{hyperref}
\usepackage{amsthm}
\usepackage{mathtools} 
\usepackage{enumitem}
\usepackage{bbm}
\usepackage{color}

\def\diag{\mathop{\rm diag}\nolimits}

\def\Ker{\mathop{\rm Ker}\nolimits}

\theoremstyle{plain}

\newtheorem*{thm*}{Theorem}

\theoremstyle{definition}

\newtheorem*{exmp*}{Example}

\usepackage[normalem]{ulem}
\usepackage{xcolor}
\usepackage {cancel}
\usepackage{CJKutf8}
\usepackage{ifthen}
\newcommand{\ADD}[1]{{#1}}

\begin{document}

\preprint{APS/123-QED}

\title{Optimal control of stochastic reaction networks with entropic control cost \\and emergence of mode-switching strategies}

\author{Shuhei A. Horiguchi}
 \altaffiliation[Also at ]{Nano Life Science Institute, Kanazawa University, Kanazawa 920-1192, Japan}
  \email{shuhei.a.horiguchi@gmail.com}
\author{Tetsuya J. Kobayashi}%
    \altaffiliation[Also at ]{Institute of Industrial Science, The University of Tokyo, Tokyo 153-8505, Japan}
    \altaffiliation[Also at ]{Universal Biology Institute, The University of Tokyo, Tokyo 113-8654, Japan}
 \email{tetsuya@mail.crmind.net}
\affiliation{%
Department of Mathematical Informatics, the Graduate school of Information Science and Technology, the University of Tokyo, Tokyo, Japan
}%

\date{\today}


\begin{abstract}
Controlling the stochastic dynamics of biological populations is a challenge that arises across various biological contexts. However, these dynamics are inherently nonlinear and involve a discrete state space, i.e., the number of molecules, cells, or organisms. Additionally, the possibility of extinction has a significant impact on both dynamics and control strategies, particularly when the population size is small. These factors hamper the direct application of conventional control theories to biological systems.
To address these challenges, we formulate the optimal control problem for stochastic population dynamics by utilizing control cost functions based on the $f$-divergence, which naturally accounts for population-specific factors. If Kullback–Leibler (KL) divergence is adopted for the cost function, the complex nonlinear Hamilton–Jacobi–Bellman equation is simplified into a linear form, facilitating efficient computation of optimal solutions.
We demonstrate the effectiveness of our approach by applying it to the control of interacting random walkers, Moran processes, and SIR models, and observe the mode-switching phenomena in the control strategies. Our approach provides new opportunities for applying control theory to a wide range of biological problems.
\end{abstract}

\maketitle


\section{Introduction}
Optimal control problems for a population of stochastically interacting particles arise in diverse fields of biology \cite{bressloff-2014-StochasticProcessesCell,qian-2021-StochasticChemicalReaction}. In intracellular chemical reactions, molecules interact stochastically and nonlinearly to generate complex dynamics, whose control is essential for medical and bioengineering applications \cite{lugagne-2017-BalancingGeneticToggle}. In cellular or animal populations, cells or organisms with different phenotypic and genetic traits interact and compete for survival and reproduction. 
Strategic control of such populations into either extinction, survival, or amplification leads to cancer therapy \cite{gatenby-2009-AdaptiveTherapy,sehl-2011-ExtinctionModelsCancer,raatz-2023-PromotingExtinctionMinimizing}, stem cell culturing \cite{mckee-2017-AdvancesChallengesStem}, manipulating gut microbiota \cite{ericsson-2015-ManipulatingGutMicrobiota}, maintenance of immunological memory \cite{woodland-2009-MigrationMaintenanceRecall,sallusto-2010-VaccinesMemoryBack}, biodiversity conservation \cite{rands-2010-BiodiversityConservationChallenges,sarkar-2006-BiodiversityConservationPlanning,kuussaari-2009-ExtinctionDebtChallenge}, and control of evolving population \cite{fischer-2015-ValueMonitoringControl,nourmohammad-2021-OptimalEvolutionaryControl}
 In human populations, control of epidemic outbreaks, spurred by individual interactions, poses a significant public health challenge \cite{piret-2021-PandemicsThroughoutHistory,aurell-2022-OptimalIncentivesMitigate,schnyder-2023-RationalSocialDistancing,avram-2024-AdvancingMathematicalEpidemiology}.

All these phenomena can be effectively formulated using the theoretical framework of reaction networks (RN), making stochastic RN theory a firm foundation for devising optimal control strategies across these areas (Hereafter, we use RN to designate this class of dynamics including chemical reactions, population dynamics, and epidemic dynamics.). 

\begin{figure*}[htp]
\centering
\includegraphics[width=18cm]{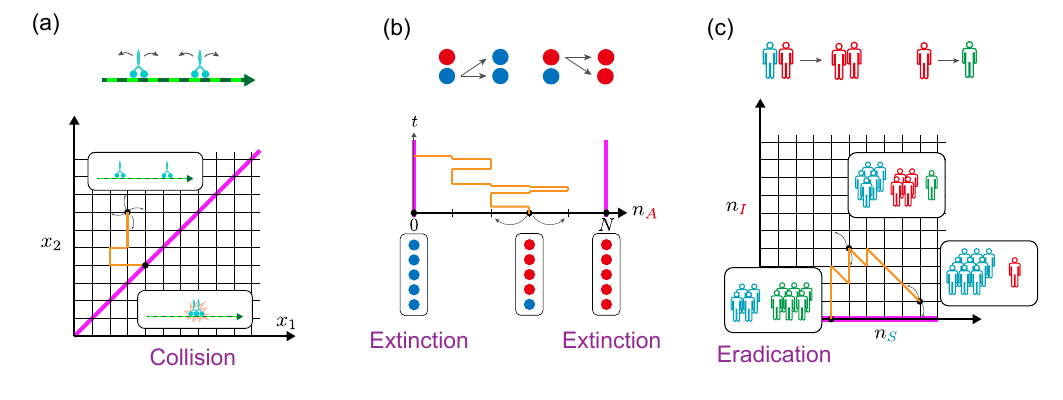}
\caption{Examples of biological phenomena described as reaction networks. (a) Movement of molecular motors on a microtubule with mutual interference. (b) Competing dynamics of populations in population genetics or ecology. (c) Spread of infectious diseases in a population. Purple thick lines in the panels represent the absorbing states in each phenomenon.}
\label{fig:intro}
\end{figure*}

Despite its broad applicability, the optimal control of RNs remains underexplored. This oversight is largely due to the unique characteristics of stochastic RNs, which deviate from the setup of conventional optimal control for diffusion processes. 
The major deviations include the inherent nonlinearity of RNs, the discrete nature of state variables (representing counts of particles like molecules or organisms), and their stochastic dynamics, which are better modeled by Markov jump processes with Poissonian randomness rather than Gaussian diffusion \cite{gardiner-2009-StochasticMethods,bressloff-2014-StochasticProcessesCell,anderson-2011-ContinuousTimeMarkov}. Additionally, the non-negativity constraint of control parameters, i.e., kinetic rate constants, requires a natural measure for cost other than conventional quadratic ones, which presume the constraint-free Euclidian parameter space. Finally, the zero count states act as intrinsic absorbing boundaries, reaching them can dramatically alter system's dynamics through the extinction of particles (Fig.~\ref{fig:intro}).

These distinct properties of RNs necessitate (1) a departure from the common control setup based on diffusion processes, such as the Linear-Quadratic-Gaussian (LQG) model, (2) an establishment of alternative theoretical framework tailored to the unique requirements of RN optimal control, and (3) its applications to biologically relevant control problems. 

In this work, we establish that all these issues can be resolved by integrating optimal control of jump processes and stochastic RN with \ADD{cost functions based on $f$-divergence~\cite{csiszar-1967-InformationtypeMeasuresDifference,morimoto-1963-MarkovProcessesHTheorem,polyanskiy-2025-InformationTheoryCoding}, which includes relative entropy or Kullback–Leibler (KL) divergence.} 
The optimal control with KL cost was originally proposed for diffusion processes in relation to the duality of control and inference \cite{fleming-1982-OptimalControlNonlinear,mitter-2003-VariationalApproachNonlinear,todorov-2008-GeneralDualityOptimal}. 
\ADD{We demonstrate that $f$-divergence, as it is designed for non-negative probability measures, can naturally accommodate the non-negativity constraints and the nature of stochastic RNs as counting processes.} 
Moreover, KL cost allows for the linearization of the nonlinear Hamilton–Jacobi–Bellman (HJB) equations through the Cole-Hopf transformation. This linearization facilitates the efficient derivation of optimal solutions in a similar manner to previously identified solvable control problems \cite{kappen-2005-PathIntegralsSymmetry,fleming-2006-ControlledMarkovProcesses,todorov-2009-EfficientComputationOptimal}.

By leveraging the sound properties of optimal control for stochastic RN with KL cost, we demonstrate the effectiveness of our theory for different biological phenomena: an optimal transport by mutually excluding molecular motors, maintenance of cellular heterogeneity in populations, and epidemic outbreak control (Fig.~\ref{fig:intro}). 
For molecular motors, we derive analytic solutions owing to the linearization. In the maintenance of heterogeneity and epidemic outbreak control, we identify mode-switching phenomena of population control that hinge on the transition in controllability of exponentially growing populations.

Thus, our new results can substantially broaden the scope of optimal control applications in stochastic chemical reactions, population dynamics, and epidemics.

\section{Optimal control of stochastic reaction systems}
\subsection{Stochastic reaction systems}

Let us consider a population of particles that evolve through stochastic events, i.e., the occurrence of reactions. Each particle is assigned one of discrete type set $X$. The number of particles of type $x \in X$ at time $t$ is denoted as $n_x(t) \in \mathbb{Z}_{\geq 0}$. The number distribution $n(t) \in \mathbb{Z}_{\geq 0}^{|X|} = \mathcal{N}$ completely characterize the state of the system at time $t$.
The number distribution changes when a reaction occurs. There are $|R|$ kinds of reactions, where $R$ is the set of reactions.
Once a reaction $r \in R$ occurs, $s^-_{r,x}\in \mathbb{Z}_{\geq 0}$ particles are consumed and $s^+_{r,x}\in \mathbb{Z}_{\geq 0}$ particles are produced. The net change in the number of particles $s_{r,x}:= s_{r,x}^+ - s_{r,x}^- \in \mathbb{Z}$ is called stoichiometric coefficient and $s_{r} = (s_{r,x})_{x\in X} \in \mathbb{Z}^{|X|}$ is stoichiometric vector for reaction $r$.
Thus when the reaction $r$ occurs at time $t$, the number distribution changes from $n(t^-)$ to
\begin{equation}
	n(t) = n(t^-) + s_r.
\end{equation}

The timing of reaction events is random, which follow an inhomogeneous Poisson process with rate $j_r(t) \geq 0$ for reaction $r$ at time $t$. It could vary over time and depends on the current number distribution $n(t)$.
We write the rate of reaction $r \in R$, or the reaction flux, as the product of the reaction rate coefficient $k_r(t) \geq 0$ and the propensity function $h_r(n)\geq 0$:
\begin{equation}\label{eq:control_affine_kinetics}
	j_r(t) = k_r(t) h_r(n(t)).
\end{equation}
When we assume the law of mass action kinetics, $k_r(t)$ is a kinetic rate constant, and the function $h_r(n)$ is given by $h_r(n) = N^{1-\sum_{x\in X} s^-_{r,x}} \prod_{x\in X} \frac{n_x!}{(n_{x}-s^{-}_{r,x})!}$, where $N$ is a constant parameter that describe the system size, e.g., the volume or total number of the particles in the system. We do not assume the mass action kinetics in the following so that our results become applicable to any propensity function. Instead, we presume only the situations that make the process well-defined up to time $T$ or up to exit time $T_{\mathrm{exit}}$. 
For example, $h_r(n)$ should be $0$ when the reaction pushes the state out of the domain, i.e., $n + s_r \not\in \mathcal{N}$.

The stochastic reaction system defined in this way can be characterized by either the counting process representation or the Markov process representation. The counting process representation  \cite{anderson-2011-ContinuousTimeMarkov} is useful for characterizing the stochastic process $(n(t))_{0\leq t \leq T}$ and simplifies the derivation of KL control cost.
Let $\xi_r(t) \in \mathbb{Z}_{\geq 0}$ denote the total number of occurrences of reaction $r$ from time $0$ to $t$. The number distribution $n(t)$ is a linear transformation of the reaction count $\xi(t)$, i.e., for any $0 \leq t \leq T$,
\begin{equation}\label{eq:reaction_count_to_numbers}
	n(t) = n(0) + \sum_{r\in R} s_r \xi_r(t).
\end{equation}
The probability law of $\xi(\cdot)$ is given by the Poisson process.
Using the time-change of independent unit Poisson processes $Y_r(t)$, the reaction count can be written as $\xi_r(t) = Y_r\left(\int_0^t k_r(\tau) h_r(n(\tau)) d\tau \right)$.

The stochastic process $n(\cdot)$ is a Markov process defined on the nonnegative and infinite integer lattice $\mathcal{N}$. The transition rate from $n\in\mathcal{N}$ to $n^\prime\in\mathcal{N}$ is given by
\begin{equation}\label{eq:controlled_transition_rate}
	\omega_{n,n^\prime}(t) = \sum_{r \in R} k_r(t) h_r(n) \delta_{n+s_r,n^\prime},
\end{equation}
where $\delta_{n,n^\prime}$ is the Kronecker delta.
Let $P_t(n^\prime) := \mathbb{E}\left[\delta_{n(t),n^\prime} | n(0) = n_0 \right]$ denote the probability of being the state $n^\prime \in \mathcal{N}$ at time $t$ given the initial state $n_0 \in \mathcal{N}$. Then, $P_t(n^\prime)$ satisfies the Kolmogorov forward equation (chemical master equation) for any $0 < t < T$ and $n\in \mathcal{N}$:
\begin{align}\label{eq:kolmogorov_forward}
	\frac{\partial}{\partial t} P_t(n) \notag
	&= \sum_{n^\prime\in\mathcal{N}\backslash\{n\}} \left[P_t(n^\prime) \omega_{n^\prime, n}(t)-P_t(n) \omega_{n,n^\prime}(t)\right] \notag \\
	&=\sum_{r\in R} k_r(t) \left[h_r(n-s_r) P_t(n-s_r)-  h_r(n) P_t(n)\right]\notag \\
	&=: \mathcal{L}_k^\dagger P_t(n).
\end{align}

\subsection{General formulation of optimal control problems for stochastic RNs}

We assume that the controller can observe the current state $n(t)$ and adjust the reaction rate coefficients $k_r(t)\geq 0$ to any desired value at any time while the function $h_r(n)$ remains unchanged, i.e., the controller can modulate only the speeds of reactions. \ADD{This is a feedback control problem so that the reaction rate coefficients $k_r(t)$ can depend on the current state $n(t)$.}
We would like to find the optimal reaction rate coefficients $(k(t))_{0 \leq t < T}$ which drive the population to take a desirable trajectory $(n(t))_{0 \leq t < T}$ while minimizing the cost of modulating the reaction rate coefficients.
Let us define the {\it utility} function $\mathcal{U}(n(\cdot)) := U_T(n(T)) + \int_0^T U_\tau(n(\tau))d\tau$ as the sum of the terminal utility function $U_T(n) \in \mathbb{R}$ and the time integral of instantaneous utility function $U_\tau(n) \in \mathbb{R}$. 
The {\it control cost} function is also defined as $\mathcal{C}(n(\cdot), k(\cdot)) := \int_0^T C(n(\tau), k(\tau))d\tau$. Then, we consider the following constrained optimization problem:
\begin{equation}
\begin{aligned}\label{eq:conjugate_value_function}
	\mathcal{E}(n_0,u) :=~&\underset{k(\cdot)}{\text{min}} ~~ \mathbb{E}_{k(\cdot)}\left[ \mathcal{C}(n(\cdot), k(\cdot))\right]\\
	&\text{s.\,t.}~~ \mathbb{E}_{k(\cdot)} \left[ \mathcal{U}(n(\cdot))\right] \geq u,
\end{aligned}
\end{equation}
or \ADD{corresponding Lagrange dual problem~\cite{boyd-2004-ConvexOptimization}}, which is given by the following unconstrained optimization:
\begin{equation}\label{eq:value_function_0}
	V(n_0, \beta) := \underset{k(\cdot)}{\text{max}} ~~ \mathbb{E}_{k(\cdot)} \left[ \beta \mathcal{U}(n(\cdot)) - \mathcal{C}(n(\cdot), k(\cdot))\right],
\end{equation}
where the expectation $\mathbb{E}_{k(\cdot)}$ is taken over trajectories $n(\cdot)$ generated under the designated reaction rate coefficients $k(\cdot)$ and initial condition $n(0)=n_0$.
A scalar $\beta$ is the Lagrange multiplier or a parameter to adjust the importance of the utility relative to the control cost.
\ADD{
Once we could find the optimal reaction rate coefficient $(k^\dagger(t))_{0 \leq t < T}$ that solves one of these problems for all $u$ or for all $\beta$ such that the solution exists, the solution to the other problem is readily obtained by the Legendre transformation:
\begin{equation}
	\mathcal{E}(n, u) = \max_\beta \left[ u \beta - V(n, \beta) \right],
\end{equation}
\begin{equation}
	V(n, \beta) = \max_u \left[ u \beta - \mathcal{E}(n, u) \right],
\end{equation}
where the relationship between $\beta$ and $u$ is given by the expected utility under the optimal control
\begin{equation}\label{eq:expected_optimal_utility}
	u = \mathcal{U}^\dagger(n, \beta) := \mathbb{E}_{k^\dagger(\cdot)}\left[ \mathcal{U}(n(\cdot))\right] = \frac{\partial}{\partial \beta} V(n,\beta).
\end{equation}
See Appendix~\ref{appendix:derivation_optimal_control}
for the derivation.
}
\ADD{Note that the formulation can be easily extended to the case with multiple constraints, e.g., $\mathbb{E}_{k(\cdot)}\left[\mathcal{U}_j(n(\cdot))|n(0)=n\right]\geq u_j$ for $j=1,\ldots,J$ or the equality constraint $\mathbb{E}_{k(\cdot)}\left[\mathcal{U}(n(\cdot))|n(0)=n\right]= u$. In the following, we focus on solving the unconstrained problem, Eq.~\eqref{eq:value_function_0}.}

We will calculate the optimal reaction rate coefficient $(k^\dagger(t))_{0 \leq t < T}$ that attains the optimum $V(n_0,\beta)$.
The standard protocol for solving the problem is to consider the value function $V_t(n, \beta)$, which is defined as the maximum of the expectation in Eq.~\eqref{eq:value_function_0} from $t$ to $T$ under the condition $n(t)=n$:
\begin{equation}\label{eq:value_function}
	V_t(n,\beta) := \max_{k(\cdot)} \mathbb{E}_{k(\cdot)}\left[
	\beta \mathcal{U}_t(n(\cdot)) - \mathcal{C}_t(n(\cdot), k(\cdot))
	| n(t) = n \right],
\end{equation}
where $\mathcal{U}_t(n(\cdot))$ and $\mathcal{C}_t(n(\cdot), k(\cdot))$ are the utility and control cost functions from time $t$ to $T$, e.g., $\mathcal{U}_t(n(\cdot)) := U_T(n(T)) + \int_t^T U_\tau(n(\tau))d\tau$.
The value function is equal to the terminal utility function at time $T$, $V_T(n, \beta) = \beta U_T(n)$, and also provides the optimum for the original problem at time $0$, $V_0(n, \beta) = V(n, \beta)$.
Thus, the reaction rate attaining Eq.~\eqref{eq:value_function} is identical to the optimal reaction rates $k^\dagger$ in the original control problem.

\ADD{
To obtain the value function, we usually apply the dynamic programming principle to derive the Hamilton–Jacobi–Bellman (HJB) equation, a system of differential equations that $V_t(n,\beta)$ satisfies: for $t \in (0,T)$,
\begin{equation}\label{eq:HJB}
	-\frac{\partial}{\partial t}V_t(n,\beta) = \beta U_t(n) + \max_{k\in \mathbb{R}_{\geq 0}^{|R|}} \left[ \mathcal{L}_k V_t(n,\beta) - C(n, k) \right],
\end{equation}
with the terminal condition $V_T(n,\beta) = \beta U_T(n)$. $\mathcal{L}_k$ is the Kolmogorov backward operator (generator), which is given by, for any scalar function $f(n)$,
\begin{equation}
	\mathcal{L}_k f(n) = \sum_{r\in R} k_r h_r(n) \overline{\nabla}_{s_r} f(n),
\end{equation}
where $\overline{\nabla}_{s_r} f(n) := f(n + s_r) - f(n)$ is the discrete directional derivative of $f$ along $s_r$. The optimal reaction rate coefficient is given by
\begin{equation}
	k^\dagger(t, n, \beta) = \arg\max_{k\in \mathbb{R}_{\geq 0}^{|R|}} \left[ \mathcal{L}_k V_t(n,\beta) - C(n, k) \right],
\end{equation}
which explicitly depends on the current time $t$, state $n$, and $\beta$.
}

\subsection{First exit optimal control problems}
We have focused on the problem with a finite and fixed terminal time $T$.
One can extend our approach to the problem with infinite time length $T\rightarrow \infty$. Since a stochastic reaction system often has absorbing states $\mathcal{N}_{\mathrm{abs}} \subset \mathcal{N}$, the stochastic process often reaches one of these absorbing states after some time $T_{\mathrm{exit}} := \min \{ t \geq 0 | n(t) \in \mathcal{N}_{\mathrm{abs}} \}$.
For example, in the birth-death processes
\begin{equation}\label{eq:birth-death_reaction}
	A \overset{k_1}{\longrightarrow} 2A,\quad
	A \overset{k_2}{\longrightarrow} \emptyset,
\end{equation}
the extinction state $n_A = 0$ is absorbing.
The extinction events are particularly important in biological problems because the extinction of some species in a chemical, organismal, or human population could alter the dynamics of the system qualitatively. Thus, controlling species into either survival or extinction has many applications, as mentioned in Introduction. 
Moreover, biological control problems may not always have a prescribed end time $T$ because the goal of control is usually to achieve something rather than to do something until $T$. Thus, the first exit control is more essential than fixed-time control for biological problems.

We can consider the optimal control problem with such a random terminal time $T_{\mathrm{exit}}$. 
The utility function is replaced with the sum of the terminal utility function $U_{\mathrm{exit}}(n)$ defined on the absorbing states $\mathcal{N}_{\mathrm{abs}}$, and the instantaneous utility function $U(n)$ defined on the non-absorbing states $\mathcal{N}_{\mathrm{non}} := \mathcal{N} \backslash \mathcal{N}_{\mathrm{abs}}$, i.e., 
\begin{equation}\label{eq:utility_function_for_first_exit}
	\mathcal{U}(n(\cdot)) = U_{\mathrm{exit}}(n(T_{\mathrm{exit}})) + \int_0^{T_{\mathrm{exit}}} U(n(\tau))d\tau.
\end{equation}

Then, the value function becomes time-independent $V(n,\beta)=V_t(n,\beta)$, and the HJB equation can be derived as
\begin{equation}\label{eq:HJB_first_exit}
	0 = \beta U(n) + \max_{k\in \mathbb{R}_{\geq 0}^{|R|}} \left[ \mathcal{L}_k V(n,\beta) - C(n, k) \right],
\end{equation}
with the boundary condition $V(n,\beta) = \beta U_{\mathrm{exit}}(n)$ for $n \in \mathcal{N}_{\mathrm{abs}}$.
Once we have the value function $V(n, \beta)$, the optimal reaction rate is given by
\begin{equation}
	k^\dagger(n, \beta) = \arg\max_{k\in \mathbb{R}_{\geq 0}^{|R|}} \left[ \mathcal{L}_k V(n,\beta) - C(n, k) \right],
\end{equation}
which depends on the current state $n$ and $\beta$ but not time $t$.

If there are no absorbing states, the terminal time and the time accumulated objective function could diverge. For such cases, the time-averaged formulation is useful:
\begin{equation}
	\underset{k(\cdot)}{\text{maximize}} \quad \lim_{T\rightarrow \infty} \frac{1}{T}\mathbb{E}_{k(\cdot)} \left[ \beta  \mathcal{U}(n(\cdot)) - \mathcal{C}(n(\cdot), k(\cdot))\right].
\end{equation}
We will not delve into details in this work but we briefly discuss the formulation in Appendix~\ref{appendix:average_cost}.

\subsection{\ADD{Choice of the control cost function}}
To obtain the value function, we usually leverage the Hamilton–Jacobi–Bellman (HJB) equation, a differential equation that $V_t(n,\beta)$ satisfies.
However, solving the HJB equation is generally intractable both analytically and even numerically because it is a nonlinear differential equation on a possibly infinite domain.
This difficulty is the major obstacle to optimal control in applications. To address the difficulty, people are conventionally forced to restrict or approximate the original problem to a linear-quadratic-Gaussian (LQG) problem, in which only the linear dynamics on a continuous Euclidean space with Gaussian noise and a quadratic control cost function can be considered. However, all these restrictions of the LQG problem conflict with the essential properties of stochastic RN control, i.e., the nonlinearity of $h_{r}(n)$, the discreteness and nonnegativity of state $n \in \mathbb{Z}_{\geq 0}^{|X|}$, the nonnegativity of the control parameter $k_{r}\geq 0$, and Poissonian nature of stochasticity. 
Several studies attempted to overcome some of these difficulties
\cite{lorch-2018-StochasticOptimalControl,theodorou-2012-StochasticOptimalControl,okumura-2017-IterativePathIntegral,briat-2022-ContinuousTimeSampledDataOptimal,briat-2022-ContinuousSampledDataControl,briat-2021-OptimalH_inftyControl}. 
Nonetheless, optimal control was not practical for biological problems described by RN.

\ADD{
In this work, we demonstrate that the difficulty can be resolved by designing the control cost function appropriately for RN.
The natural way to measure control costs would vary depending on the context. However, we find some basic assumptions that may be applied to the control cost function in most practical situations.
Noting the similarity of these assumptions to the properties of the divergence, which is the dissimilarity measure between a pair of probability measures or nonnegative measures in information theory, we propose using the divergence to formulate the control cost.
}

\ADD{
First, the control cost $C$ should be a function of the increment in reaction counts $\xi_r$ or the reaction flux $j_r=\frac{d}{dt}\mathbb{E}[\xi_r]=k_r h_r(n)$, by which the control cost function depends on the state $n$ and control input $k$.
This is a natural assumption, for instance, when the control of the reaction network is implemented by intermittent interventions activated at every chance of a reaction event.
}

\ADD{
Second, the instantaneous control cost function $C(n, k)$ should have a unique minimizer $k^0(n) = \arg\min_k C(n, k)$ for each $n\in \mathcal{N}$. Without loss of generality, the minimum control cost can be set to be zero, $C(n, k^0(n)) = 0$.
Note that in conventional control problems such as LQG problems, the control cost function is given by a quadratic function, which is minimized when the control input is zero, $k^0_r = 0$. However, the non-zero (positive) uncontrolled input $k^0_r > 0$ is common in controlling reaction networks, for instance, when the controller can suppress reactions that otherwise lead to undesirable outcomes.
}

\begin{figure*}[htp]
\centering
\includegraphics[width=15cm]{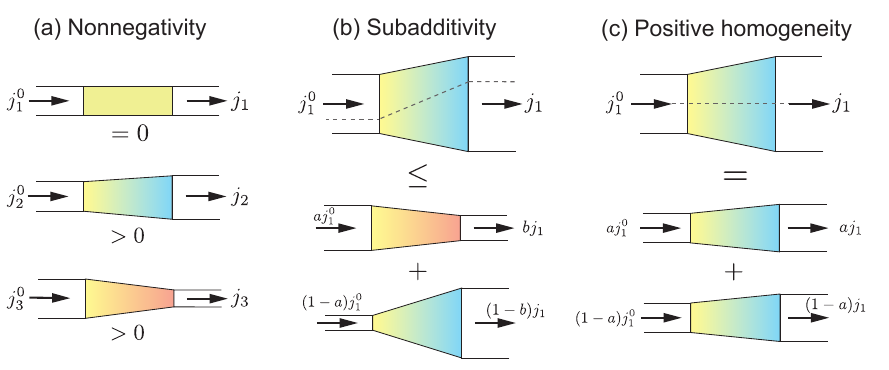}
\caption{\ADD{Diagram showing the properties of the control cost. 
The control of reaction fluxes is visualized as the control of flow speed in channels. $j_r^0$ and $j_r$ are the uncontrolled (input) and the controlled (output) reaction fluxes, respectively. $a, b \in [0, 1]$ are the ratios of dividing the input and output fluxes.}}
\label{fig:control_cost}
\end{figure*}

\ADD{
The dynamics with vanishing control cost can be interpreted as the {\it uncontrolled dynamics}. Thus, the control cost function $C(n, k)$ measures the deviation of the current control input $k$ from the uncontrolled one $k^0(n)$.
Combined with the first assumption, the control cost function can be interpreted as the control of reaction flux from the uncontrolled one $j_r^0(n) := k^0_r(n) h_r(n)$ to the controlled one $j_r$~(Fig. \ref{fig:control_cost}~(a)). 
Noting that the reaction fluxes $j$ and $j^0$ can be seen as nonnegative measures on the set of reactions $R$, the divergence in information theory may serve as a control cost function.
}

\ADD{
Specifically, we propose to use $f$-diverngence-like function
\begin{equation}\label{eq:f_divergence_control_cost}
	C(n, k) = \sum_{r\in R} j^0_r f_r\left(\frac{j_r}{j^0_r}\right)
	=\sum_{r\in R} h_r(n) k^0_r f_r\left(\frac{k_r}{k^0_r}\right),
\end{equation}
where $f_r:[0, +\infty)\rightarrow [0, +\infty]$ is a convex function satisfying $f_r(1)=0$ and $f_r^\prime(1)=0$ for $r\in R$. Note that the standard definition of the $f$-divergence in the information theory~\cite{csiszar-1967-InformationtypeMeasuresDifference,polyanskiy-2025-InformationTheoryCoding} adopts a reaction-independent function $f_r=f$ in Eq.~\eqref{eq:f_divergence_control_cost} to ensure permutation invariance, but we extend it to allow reaction dependence.
For example, using a quadratic function $f_r(x)=\frac{1}{2}(x-1)^2$, we obtain $\chi^2$-divergence-like control cost function
\begin{equation}\label{eq:quadratic_control_cost}
	C_{quad}(n, k) = \sum_{r\in R}  h_r(n) \frac{\left(k_r - k^0_r\right)^2}{2k_r^0}.
\end{equation}
Quadratic cost functions weighted by the propensity $h_r(n)$ have been used in quantifying the biological cost of cellular reactions \cite{horiguchi-2023-CellularGradientFlow}.
}

\ADD{
In fact, $f$-divergence is the only class of divergences with (1) decomposability, (2) subadditivity, and (3) positive homogeneity, all of which are reasonable assumptions as the control cost.
The decomposability means that the total control cost is a sum of the control cost for each reaction $r\in R$ so that the control cost function can be written as $C = \sum_{r\in R} D_r(j_r, j^0_r)$, which implies the independence of the control mechanisms or flow-controlled channels for each reaction. 
The subadditivity and positive homogeneity mean that
\begin{equation}
	D_r(j_r, j^0_r) \leq D_r(a j_r, b j^0_r) + D_r((1-a)j_r, (1-b)j^0_r),
\end{equation}
\begin{equation}
	D_r(\rho j_r, \rho j^0_r) = \rho D_r(j_r, j^0_r),
\end{equation}
holds, respectively, for any $j_r, j^0_r,\rho \geq 0$ and $a,b \in [0, 1]$.
The subadditivity implies that the control cost increases if the input and output fluxes of a channel are divided into those of two independent channels~(Fig.~\ref{fig:control_cost}~(b)).
However, positive homogeneity implies that if the division ratios for input and output fluxes are the same, $a=b$, the two channels are identical, and control can be done efficiently without increasing the control cost~(Fig.~\ref{fig:control_cost}~(c)).
In Appendix~\ref{appendix:f_divergence}, we prove that the class of control cost functions satisfying the above assumptions is exactly $f$-divergence-like functions in Eq.~\eqref{eq:f_divergence_control_cost}. 
}

\ADD{
Furthermore, provided an admissible control exists, we can guarantee the existence of a unique optimal reaction rate coefficient by imposing 1-coercivity  and strict convexity on the function $f_r$.
The 1-coercivity, i.e.,
\begin{equation}
	\lim_{x \rightarrow +\infty}\frac{f_r(x)}{x} = +\infty,
\end{equation}
and the strict convexity, i.e.,
\begin{equation}
	f_r(a x + (1-a) x^\prime) < a f_r(x) + (1-a) f_r(x^\prime)
\end{equation}
for any $a \in (0,1)$ and $x, x^\prime \geq 0$, 
guarantee the existence and uniqueness of the optimum, respectively.
}

\ADD{
Using the $f$-divergence-like control cost function, the HJB Eq.~\eqref{eq:HJB} can be rewritten as
\begin{equation}\label{eq:HJB_conjugate_f}
	-\frac{\partial}{\partial t}V_t(n,\beta) = \beta U_t(n) + \sum_{r\in R}k^0_r h_r(n) f_r^*(\overline{\nabla}_{s_r} V_t(n, \beta)),
\end{equation}
where $f_r^*(y) := \max_x [xy - f_r(x)]$ is the Legendre transform of $f_r(x)$.
The optimal reaction rate coefficient $k^\dagger_r(t, n, \beta)$ is given by the following time-dependent state-feedback controller
\begin{equation}\label{eq:optimal_control_conjugate_f}
	k^\dagger_r(t, n, \beta) = k^0_r \partial f_r^*(\overline{\nabla}_{s_r} V_t(n, \beta)).
\end{equation}
As for the first exit problems, HJB Eq.~\eqref{eq:HJB_first_exit} becomes
\begin{equation}\label{eq:HJB_first_exit_f}
	0 = \beta U(n) + \sum_{r\in R} k_r^0 h_r(n) f_r^*(\overline{\nabla}_{s_r} V(n,\beta)),
\end{equation}
and the optimal reaction rate is given by
\begin{equation}
	k^\dagger_r(n, \beta) = k^0_r \partial f_r^*(\overline{\nabla}_{s_r} V(n, \beta)).
\end{equation}
}

\ADD{
In the case of quadratic control cost, the Legendre transform of $f_r(x)=\frac{1}{2}(x-1)^2$ is given by 
\begin{equation}
	f^*_r(y) = \frac{1}{2} \left([y + 1]_+^2 - 1\right),\quad
	\partial f^*_r(y)= [y + 1]_+,
\end{equation}
where $[a]_+ = \max\{a, 0\}$ is the positive part of a scalar $a$.
The positive part operation $[\cdot]_+$ comes from the non-negativity constraint on $k$.
}

\ADD{
Therefore, the $f$-divergence-like control cost functions are intuitive and mathematically convenient in formulating stochastic RN control problems.
}

\subsection{Optimal control with KL cost}
\ADD{
Among a variety of $f$-divergence-like control cost functions, we found that the following Kullback–Leibler (KL) divergence-like control cost function is exceptionally useful:
\begin{equation}\label{eq:KL_cost_function}
	C_{KL}(n, k) = \sum_{r \in R} h_r(n) \left(k_r \log \frac{k_r}{k_r^0} - k_r + k_r^0\right),
\end{equation}
which is given by setting $f_r(x)=x\log x - x + 1$ in Eq.~\eqref{eq:f_divergence_control_cost}.
The value at $x=0$ is defined by the limit $f_r(0) := \lim_{x\rightarrow 0}f_r(x)=1$.
}

The optimal reaction coefficient for $r\in R$ at time $t$ is given by
\begin{equation}\label{eq:optimal_reaction_rate}
	k^\dagger_r(t, n, \beta) 
	= k^0_r \exp \left( \overline{\nabla}_{s_r} V_t(n, \beta) \right),
\end{equation}
(see Appendix~\ref{appendix:derivation_optimal_control} for the derivation).
As the slope of $C_{KL}$ goes to infinity, $\frac{\partial}{\partial k_r}C_{KL}(n, k)\rightarrow -\infty$ for $k_r\rightarrow 0$, the nonnegativity constraint over $k$ is automatically satisfied. In fact, the exponential function in Eq.~\eqref{eq:optimal_reaction_rate} ensures that the optimal coefficient $k^\dagger_r$ is non-negative. 

For the KL cost function, the HJB equation becomes
\begin{equation}\label{eq:HJB_for_V}
\begin{aligned}
	-\frac{\partial}{\partial t}V_t(n, \beta) &= \beta U_t(n) \\
	&+ \sum_{r \in R} k^0_r h_r(n) \left( \exp \left( \overline{\nabla}_{s_r} V_t(n, \beta) \right) - 1 \right),
\end{aligned}
\end{equation}
which is yet a nonlinear differential equation. However, by the Cole–Hopf transformation (logarithmic transformation), $Z_t(n,\beta) := \exp (V_t(n, \beta))$, the HJB equation is linearized as
\begin{equation}\label{eq:HJB_for_Z}
\begin{aligned}
	-\frac{\partial}{\partial t} Z_t(n, \beta) 
	&= \beta U_t(n) Z_t(n, \beta) + \sum_{r \in R} k^0_r h_r(n) \overline{\nabla}_{s_r} Z_t(n,\beta)\\
	&=: \beta U_t(n) Z_t(n,\beta) + \mathcal{L}_{k^0} Z_t(n, \beta),
\end{aligned}
\end{equation}
with the terminal condition $Z_T(n,\beta) = \exp(\beta U_T(n))$. 

Although it is still an ordinary differential equation on a possibly infinite domain, the linearity allows us to calculate the value function efficiently.
Note that the second term on the right-hand side is the Kolmogorov backward operator $\mathcal{L}_{k}$, which is the adjoint of the Kolmogorov forward operator $\mathcal{L}^\dagger_k$ in the chemical master equation (Eq.~\eqref{eq:kolmogorov_forward}).
Moreover, the special form of Eq.~\eqref{eq:HJB_for_Z} allows us to have the following probabilistic representation for $Z_t(n,\beta)$:
\begin{equation}\label{eq:probabilistic_rep_expvalue_reaction}
	Z_t(n, \beta) = \mathbb{E}_{k^0} \left[ \exp\left( \beta \mathcal{U}_{t}(n(\cdot)) \right) | n(t) = n\right],
\end{equation}
which is called the Feynmann-Kac formula, cf., Appendix~1.~Prop.~7.1 of \cite{kipnis-1999-ScalingLimitsInteracting}.
Thus, the value function is given by
\begin{equation}\label{eq:probabilistic_rep_value_reaction}
	V_t(n,\beta) = \log\mathbb{E}_{k^0} \left[ \exp\left( \beta \mathcal{U}_{t}(n(\cdot)) \right) | n(t) = n\right].
\end{equation}
This representation enables us to compute the value function by evaluating the expectation of the utility function with respect to the uncontrolled reaction rate coefficient $k^0$.
As demonstrated in the following sections, for simple reaction systems, we can obtain the analytical expression of the value function. 
Even if an analytical expression is inaccessible, efficient Monte Carlo sampling techniques can be used to estimate the expectation.
It is worth noting that this representation of the value function can be evaluated in a time-forward manner, whereas the usual optimal control problem requires the time-backward calculation of the HJB equation due to the dynamic programming principle.

The linearization of the HJB equation and efficient computation of the optimal control problems for certain control cost functions was reported by Kappen \cite{kappen-2005-PathIntegralsSymmetry} for diffusion processes and by Todorov \cite{todorov-2009-EfficientComputationOptimal} for discrete-time Markov chains. As Theodorou and Todorov \cite{theodorou-2012-RelativeEntropyFree} discussed, the linearization of the HJB equation is possible if the control cost function is given by KL divergence between path measures.  In this case, the optimal control problem is related to an estimation (filtering) problem. Similar mathematical properties had been found in relation to the duality of control and inference \cite{fleming-1982-OptimalControlNonlinear,mitter-2003-VariationalApproachNonlinear,todorov-2008-GeneralDualityOptimal}.
Similar properties for Markov jump processes are identified in recent studies \cite{gao-2023-OptimalControlFormulation,gueant-2020-OptimalControlFinite,jaimungal-2024-MinimalKullbackLeibler} as well as one of the earliest studies by Fleming \cite{fleming-1982-OptimalControlNonlinear}.
In this paper, we develop the optimal control framework for stochastic RN inspired by these previous studies.
In Appendix~\ref{appendix:KL_path_measure}, we elaborate on the path measure perspective and why the KL cost function in Eq.~\eqref{eq:KL_cost_function} works. See Appendix~\ref{appendix:previous_studies} for further comparison with existing literature.

The same trick applies to the first exit problems.
The Cole–Hopf transformation $Z(n, \beta) := \exp(\beta V(n, \beta))$ yields
\begin{equation}\label{eq:fisrt_exit_linear_HJB_equation}
	0 = \beta U(n)Z(n, \beta) + \mathcal{L}_{k^0} Z(n, \beta),
\end{equation}
for $n\in \mathcal{N}_{\mathrm{non}}$, and $Z(n, \beta) = \exp\left(\beta U_{\mathrm{exit}}(n, \beta)\right)$ for $n\in \mathcal{N}_{\mathrm{abs}}$.
\ADD{The equations can be written in the matrix-vector form:
\begin{equation}\label{eq:first_exit_linear_algebra}
	\bm{0} = \left(\beta \diag\bm{U} + \Omega^0_{\mathrm{non},\mathrm{non}}\right)\bm{Z}_{\mathrm{non}}
	+ \Omega^0_{\mathrm{non},\mathrm{abs}} \bm{Z}_{\mathrm{abs}},
\end{equation}
where $\bm{U} = \{ U(n)\}$ is a $|\mathcal{N}_{\mathrm{non}}|$-dimensional vector and the matrices $\Omega^0_{\mathrm{non},\mathrm{non}} \in \mathbb{R}^{|\mathcal{N}_{\mathrm{non}}|\times|\mathcal{N}_{\mathrm{non}}|}$ and $\Omega^0_{\mathrm{non},\mathrm{abs}}\in\mathbb{R}^{|\mathcal{N}_{\mathrm{non}}|\times|\mathcal{N}_{\mathrm{abs}}|}$ are given by the the uncontrolled transition rate $\omega^0_{n,n^\prime}:=\sum_{r\in R} k^0_r h_r(n) \delta_{n+s_r, n^\prime}$ as in Eq.~\eqref{eq:controlled_transition_rate}. 
From the boundary condition, $\bm{Z}_{\mathrm{abs}}$ is known. We can solve the above equation for $\bm{Z}_{\mathrm{non}}$.}
Thus, the first exit optimal control problem is reduced to a linear algebraic problem. 
Similar results have been obtained for discrete-time Markov chains \cite{todorov-2009-EfficientComputationOptimal}.
The probabilistic representations in Eqs.~\eqref{eq:probabilistic_rep_expvalue_reaction}, \eqref{eq:probabilistic_rep_value_reaction} are also applicable.

If there are no absorbing states, the terminal time and the time accumulated objective function could diverge. For such cases, the time-averaged formulation is useful:
\begin{equation}
	\underset{k(\cdot)}{\text{maximize}} \quad \lim_{T\rightarrow \infty} \frac{1}{T}\mathbb{E}_{k(\cdot)} \left[ \beta  \mathcal{U}(n(\cdot)) - \mathcal{C}(n(\cdot), k(\cdot))\right].
\end{equation}
Via the Cole–Hopf transformation, the optimal solution could be cast into an eigenvalue problem 
(see Appendix~\ref{appendix:average_cost}).

\subsection{Controlling a random walker}

\begin{figure*}[htp]
\centering
\includegraphics[width=16cm]{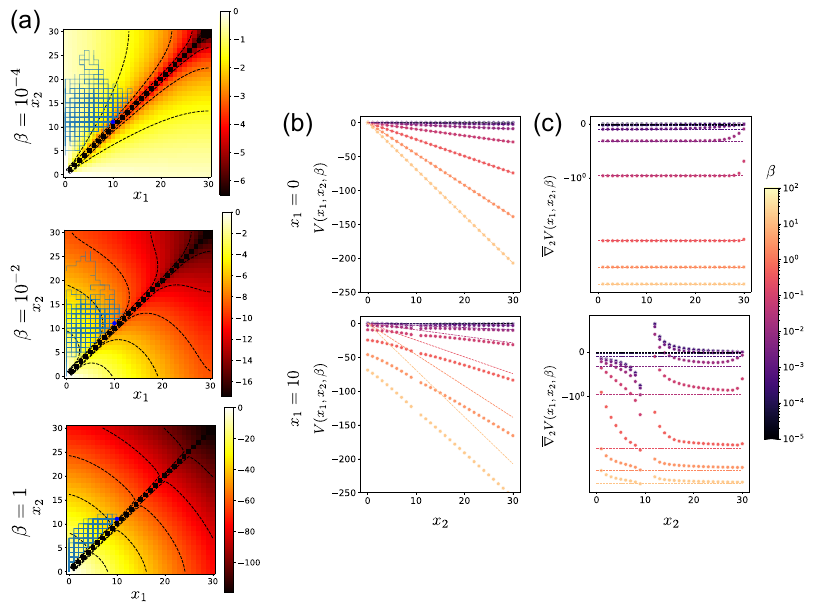}
\caption{The results of the minimum exit time problem for interacting random walkers on one-dimensional space. 
(a) The value function $V(x_1, x_2, \beta)$ plotted by color in the state space $(x_1, x_2)$ with different $\beta$. The dashed curves are its contours.
The blue zigzag lines are the trajectories of $100$ independent simulations starting from $(10, 11)$. 
(b) The value function $V(x_1, x_2, \beta)$ for fixed $x_1 = 0$ (the upper panel) and $x_1 = 10$ (the lower panel) with different $\beta$ indicated by the dots. 
The dashed lines are calculated with Eq.~\eqref{eq:value_minimum_time_random_walk}. The color code represents the value of $\beta$.
(c) The directional derivative $\overline{\nabla}_2 V(x_1, x_2, \beta) = V(x_1, x_2+1, \beta) - V(x_1, x_2, \beta)$ of the value function in $x_2$ direction for fixed $x_1 = 0$ and $x_1 = 10$. The dashed lines correspond to Eq.~\eqref{eq:value_minimum_time_random_walk}. The format of the panels is the same as in (b).
The parameters are $M=31$, $N=2$, $k^0=0.1$.}
\label{fig:interacting_random_walk_first_exit}
\end{figure*}

We demonstrate the effectiveness of our method by applying it to several control problems.
The first example is the jump processes on discrete states $X$. We consider $N$ particles walking on a directed graph $(X, R)$, where the set of directed edges $R$ represents the allowed transitions between types $X$. For any $x,y\in X$, $(x,y) \in R$ implies that the particle can jump from $x$ to $y$. The stoichiometric coefficients are given by $s_{(x,y),z} = \delta_{y,z} - \delta_{x,z}$ and the kinetics is the mass action type $h_{(x,y)}(n) = n_x$ for $x,y,z \in X$.

For $X = \mathbb{Z}$ and $N=1$, the process is reduced to a simple one-dimensional random walk by a single walker. 
Such a process has been used as a simple model of intracellular transport of macromolecules along intracellular filaments \cite{xie-2020-TheoreticalAnalysisDynamics}.
Molecular motors such as dynein and kinesin consume energy and move in one direction. Since the step length is fixed, the position of a molecular motor on a filament can be modeled as a 1-dimensional discrete grid.
For this case, we can have analytical solutions for some optimal control problems owing to the sound properties of KL control.

Consider the situation where a particle is required to reach a goal $x^* \in \mathbb{Z}$ as soon as possible. The position of the particle at time $t$ is denoted as $\hat{x}(n(t)) \in X$. 
We can formulate the situation as the following minimum exit time problem:\ADD{
\begin{equation}\label{eq:random_walker_first_exit_problem}
	\underset{k(\cdot)}{\text{minimize}} \quad  \mathbb{E}_{k(\cdot)}\left[ \beta T_{\mathrm{exit}} + \int_0^{T_{\mathrm{exit}}} C(n(t), k(t))dt \right],
\end{equation}}
where the absorbing states are given by $\mathcal{N}_{\mathrm{abs}} = \{ n\in\mathcal{N}|\hat{x}(n)=x^*\}$.
\ADD{
By setting $U_{\mathrm{exit}}(n) = 0$ and $U(n)=-1$ in Eq.~\eqref{eq:utility_function_for_first_exit}, we have the following maximization problem which is equivalent to Eq.~\eqref{eq:random_walker_first_exit_problem} having the same form as Eq.~\eqref{eq:value_function_0}:
\begin{equation}
	\underset{k(\cdot)}{\text{maximize}} \quad  \mathbb{E}_{k(\cdot)}\left[ -\beta T_{\mathrm{exit}} - \mathcal{C}(n(\cdot), k(\cdot)) \right].
\end{equation}
}
If we adopt the KL control cost function, Eq.~\eqref{eq:probabilistic_rep_value_reaction} shows that the value function $V(n, \beta) = V(\hat{x}(n),\beta)$ is equal to the cumulant-generating function of the exit time with parameter $-\beta$ and given by
\begin{equation}
	V(x, \beta) = \log \mathbb{E}_{k^0}[\exp(-\beta T_{\mathrm{exit}})|x(0)=x].
\end{equation}
Assuming the symmetric uncontrolled transition rates $k^0_r = \kappa$ for all $r\in R$ and using the analytical expression of the cumulant-generating function~\cite{feller-1966-InfinitelyDivisibleDistributions}, we obtain the value function analytically:
\begin{equation}\label{eq:value_minimum_time_random_walk}
	V(x, \beta) = -\gamma(\beta)|x^* - x|,
\end{equation}
where a scalar $\gamma(\beta) \geq 0$ is defined as
\begin{equation}
	\gamma(\beta) := - \log\left(1 + \frac{\beta}{2\kappa} - \sqrt{\left(1+\frac{\beta}{2\kappa}\right)^2 - 1}\right).
\end{equation}
The conjugate value function can be calculated for $x^* \neq x$ and $u<0$ as follows
\begin{equation}\label{eq:conjugate_value_exit_time_random_walk}
	\mathcal{E}(x, u) = 2\kappa |u| - \sqrt{\Delta^2  + 4\kappa^2 u^2} - \Delta \sinh^{-1}\left(\frac{\Delta}{2\kappa |u|}\right),
\end{equation}
where $\Delta := x^* - x$.
Since the value function is linear in $|x^*-x|$, we have the piecewise constant optimal transition rates \ADD{by using Eqs.~\eqref{eq:optimal_reaction_rate} and \eqref{eq:value_minimum_time_random_walk}}:
\begin{equation}
\begin{aligned}
	k^\dagger_{(x,x+1)}(n, \beta) &= \begin{cases}
		\kappa e^{\gamma(\beta)} & \text{if}~x<x^*,\\
		\kappa e^{-\gamma(\beta)} & \text{if}~x > x^*,
	\end{cases}\\
	k^\dagger_{(x,x-1)}(n,\beta) &= \begin{cases}
		\kappa e^{-\gamma(\beta)} & \text{if}~x < x^*,\\
		\kappa e^{\gamma(\beta)} & \text{if}~x>x^*.
	\end{cases}
\end{aligned}
\end{equation}
When $x < x^*$, the transition rate $k^\dagger_{(x,x+1)}$ in the positive direction is higher than the uncontrolled one $\kappa$, while the transition rate $k^\dagger_{(x,x-1)}$ in the negative direction is lower than $\kappa$.
We can also obtain the analytic solution for the standard control problem to maximize the average speed of the walker (see Appendix~\ref{appendix:random_walk_finite_horizon}).

\subsection{Controlling interacting random walkers}
Next, we consider the case where $N=2$ and $X=\mathbb{Z}$.
Let us consider the minimum time problem with exclusion interaction, i.e., all the particles should reach the goal $x^* \in X$ as soon as possible while they are not allowed to occupy the same site.
This is a model of two molecular motors moving on the same filament on which they cannot overtake (Fig.~\ref{fig:intro}~(a)).
Non-colliding random walks and diffusion processes have been studied persistently \cite{karlin-1959-CoincidenceProbabilities,fisher-1984-WalksWallsWetting,katori-2007-NoncollidingBrownianMotion}. 

Let us denote the position of the left particle as $\hat{x}_1(n)$ and the right particle as $\hat{x}_2(n)$, i.e., $\hat{x}_1(n) < \hat{x}_2(n)$. Due to the exclusive interaction, the first particle cannot overtake the second. Thus, the particles are identifiable for all $t \geq 0$.
The absorbing states are given by
\begin{equation}
	\mathcal{N}_{\mathrm{abs}} = \left\{ n \in \mathcal{N} ~| \hat{x}_1(n) = \hat{x}_2(n) \right\}.
\end{equation}

Among the abosrbing states above, we are interested in the single state $n^*\in \mathcal{N}_{\mathrm{abs}}$ such that the particles reach the goal $\hat{x}_1(n^*)=\hat{x}_2(n^*)=x^*$.
We can formulate the problem as a first exit problem with the instantaneous utility $U(n)=-1$, and the terminal utility 
\begin{equation}\label{eq:exit_utility_interacting_walkers}
	U_{\mathrm{exit}}(n) = \begin{cases}
		0 & \text{if}~n = n^*,\\
		-\infty & \text{otherwise}.
	\end{cases}
\end{equation}

Assuming that each particle stops when it reaches the goal, the time-integrated instantaneous utility is given by
\begin{equation}
	T_{\mathrm{exit}}
	= \max\{T^1_{\mathrm{exit}}, T^2_{\mathrm{exit}}\},
\end{equation}
where $T^i_{\mathrm{exit}} := \inf\{ t\geq 0 | \hat{x}_i(n(t)) = x^*\}$ is the first exit time of the $i$-th particle ($i=1,2$).
Then, the value function $V(n, \beta) =: V(\hat{x}_1(n), \hat{x}_2(n), \beta)$ can be calculated as
\begin{equation}
\begin{aligned}
	V(x_1, x_2, \beta) &= \mathbb{E}_{k^0}\left[\exp(-\beta T_{\mathrm{exit}}) | n(0) = n, n(T_{\mathrm{exit}})=n^* \right]\\
	&= \mathbb{E}_{k^0}\left[\mathbb{E}_{k^0}\left[\exp(-\beta \max\{T^1_{\mathrm{exit}}, T^2_{\mathrm{exit}}\}) \right.\right. \\
	&\quad\quad\big|\left.\left. x_1(0) = x_1, x_1(T^1_{\mathrm{exit}})=x^*\right]\right.\\
	&\quad\quad\big|\left. x_2(0)=x_2, x_2(T^2_{\mathrm{exit}})=x^*\right]\\
	&\leq \min\{ V(x_1, \beta), V(x_2, \beta)\}\\
	&= -\gamma(\beta) \max\{  |x^* - x_1|, |x^* - x_2|\},
\end{aligned}
\end{equation}
where $V(x_i, \beta)$ is the value function in Eq.~\eqref{eq:value_minimum_time_random_walk} for a single random walker.
When the first particle is close to the goal $x_1 \approx x^*$ while the second particle is still away from it, $|x_2 - x^*| \gg 1$, the upper bound become tight and we have
\begin{equation}\label{eq:approx_value_interacting_random_walk}
	V(x_1, x_2, \beta) \approx -\gamma(\beta) |x^* - x_2|.
\end{equation}
Using the upper bound, the conjugate value function $\mathcal{E}(x_1, x_2, u)$ satisfies
\begin{equation}\label{eq:lower_bound_of_random_walkers_conjugate_value}
	\mathcal{E}(x_1, x_2, u) \geq \max\{ \mathcal{E}(x_1, u), \mathcal{E}(x_2, u) \},
\end{equation}
where $\mathcal{E}(x_i, u)$ is the conjugate value function in Eq.~\eqref{eq:conjugate_value_exit_time_random_walk}.

Analytical estimates are compared with numerical solutions in Fig.~\ref{fig:interacting_random_walk_first_exit}. 
Numerical solutions are obtained by solving the linear equation in Eq.~\eqref{eq:fisrt_exit_linear_HJB_equation} where the infinite space $\mathbb{Z}$ is truncated to the finite interval $X = \{ 0, 1, \ldots, M - 1\}$. The goal position is set to $x^*=0$.

When the first particle has already reached the goal $x_1 = 0 = x^*$, the problem is equivalent to the minimum time problem for the second particle only.
Then, the value function satisfies $V(x^*, x_2, \beta) = -\gamma(\beta) |x^* - x_2|$, which is consistent with Eq.~\eqref{eq:approx_value_interacting_random_walk} (Fig.~\ref{fig:interacting_random_walk_first_exit}~(b), upper panel).
Then, the derivative of $V(x^*, x_2, \beta)$ is constant (Fig.~\ref{fig:interacting_random_walk_first_exit}~(c), upper panel), which results in the constant optimal control.

When the first particle is at $x_1 = 10$, the value $V(10, x_2, \beta)$ as a function of $x_2$ has a gentle slope than $\gamma(\beta)$ to avoid collision with the first particle (Fig.~\ref{fig:interacting_random_walk_first_exit}~(b), lower panel). 
Especially when $\beta$ is small, collision avoidance is more important than early arrival. 
In this case, the derivative of the value can be positive, and the second particle moves away from the first particle, as seen in Fig.~\ref{fig:interacting_random_walk_first_exit}~(c) lower panel. 
On the other hand, when $\beta$ is large, the exit time becomes more important than the collision. Thus, the derivative is always negative except at the collision point, i.e., at $x_2 = 10$.

\subsection{Controlling survival in birth and death processes}

\begin{figure*}[htp]
\centering
\includegraphics[width=16cm]{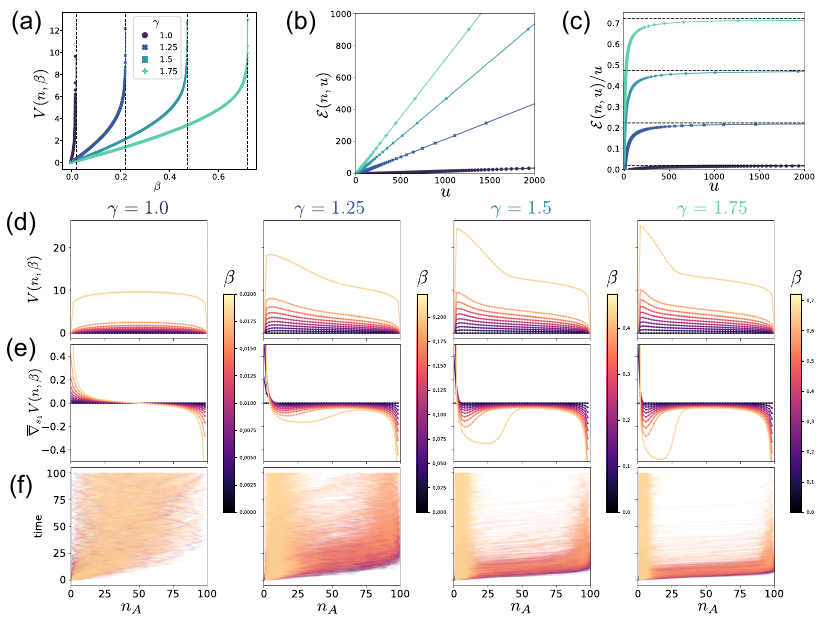}
\caption{The results of the maximum exit time problem for Moran processes with $N=100$. (a) Dependence of the value function $V(n, \beta)$ on $\beta$ at $n_A=N/2$ with varying $\gamma$, the selective advantage of $A$ over $B$. $\gamma=1$ is the neutral case.  Both the linear algebraic solution (colored points) and the analytical solution (colored curves) show divergence at $\beta=\beta_c$ (black dashed lines). (b) The relationship $\mathcal{E}(n, u)$ between the expected first exit time $u$ and the control cost starting from the state $n_A=N/2$. The format of the plot is the same as in (a). (c) The relationship $\mathcal{E}(n, u)/u$ between the expected first exit time and the control cost rate starting from the state $n_A=N/2$. The black dashed lines represent $\beta_c$ for each $\gamma$. The format of the plot is the same as in (a).
(d, e) Dependence of the value function $V(n, \beta)$ (d) and its derivative $\overline{\nabla}_{s_1}V(n, \beta)$ (e) on the initial state $n_A$ for different values of $\beta$ and $\gamma$. In each panel, the values of $\beta$ are sampled at equal intervals between $0$ to $\beta_c$ and color-coded.
(f) The optimally controlled stochastic trajectories from time $0$ to $T=100$ for different $\beta$ and $\gamma$. 
In each panel, there are $100$ trajectories with initial conditions $n_A(0)=1$ for different values of $\beta$, which are color-coded. 
The parameters are $k_1^0=\gamma, k_2^0=1.0$.}
\label{fig:Moran_first_exit}
\end{figure*}

\begin{figure*}[htp]
\centering
\includegraphics[width=18cm]{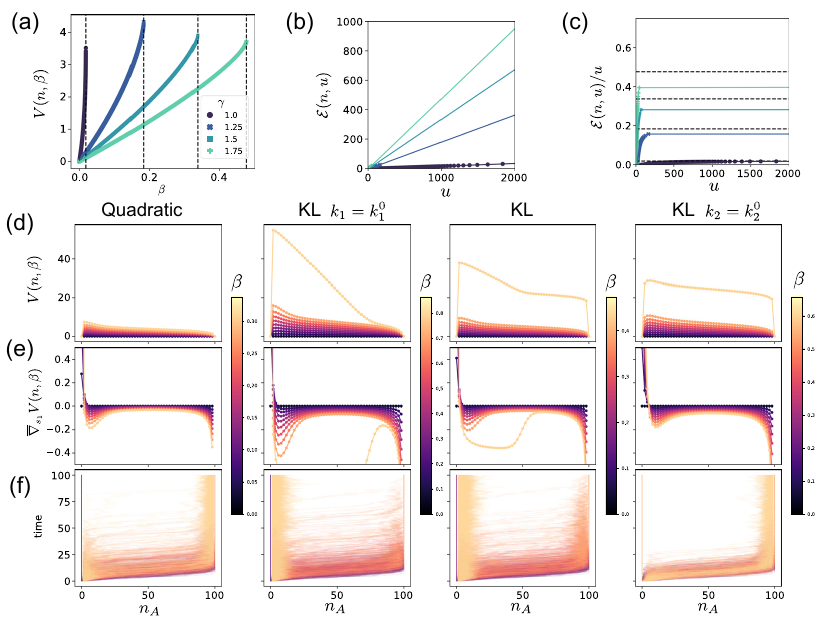}
\caption{\ADD{Comparison of maximum exit time solution for Moran processes with different control cost functions. (a–c) The value functions with the quadratic control cost function $C_{quad} $ are shown as in Fig.~\ref{fig:Moran_first_exit}~(a-c). (d–f) The value functions and optimally controlled trajectories with the quadratic $C_{quad}$, weighted KL $C_{wKL}$ ($w_1=\infty, w_2=1$), KL $C_{KL}$, and weighted KL $C_{wKL}$ ($w_1=1, w_2 = \infty$) control cost functions.
The parameters are $N=100, k_1^0=1.5, k_2^0=1.0$.}
}
\label{fig:Moran_other_cost}
\end{figure*}

In the control of birth and death processes of cells or organisms without immigration, the extinction state $n=0$ is a natural absorbing state. 
In the context of biodiversity conservation, one has to avoid extinction.
Furthermore, a single species should not dominate the ecosystem.
To address this population control problem,
let us consider the following birth and death reactions involving two species $A$ and $B$ (Fig.~\ref{fig:intro} (b)):
\begin{equation}
\begin{aligned}
	A \rightarrow 2A, \quad A \rightarrow \emptyset,\\
	B \rightarrow 2B, \quad B \rightarrow \emptyset.
\end{aligned}
\end{equation}
Assume that every dying individual is replaced by a duplicated individual of either $A$ or $B$, and that the size $N$ of the population is constant.
This is the continuous-time Moran process studied in population genetics \cite{moran-1958-RATEAPPROACHHOMOZYGOSITY,karlin-1967-NumberMutantsMaintained,houchmandzadeh-2010-AlternativeDiffusionEquation}.
Since birth and death reactions occur simultaneously, we can summarize the four reactions into two:
\begin{equation}
	A + B \overset{k_1}{\rightarrow} 2A,\quad A+B \overset{k_2}{\rightarrow} 2B.
\end{equation}
where the propensity functions are given by $h_1(n) = h_2(n) = n_A n_B / N$.
We denote the ratio between the rate coefficient by $\gamma := k_1^0 / k_2^0$, which means that type $A$ has a selective advantage over type $B$ when $\gamma > 1$ and vice versa. The case with $\gamma = 1$ is the neutral situation. 
In any case, the system eventually reaches one of two extinction boundaries $(n_A, n_B)=(N, 0)$ or $(0, N)$ due to the random drift.
We formulate the maintenance of the population coexistence as the maximization of the exit time problem from coexisting states:
\ADD{
\begin{equation}
	\underset{k(\cdot)}{\text{maximize}} \quad  \mathbb{E}_{k(\cdot)}\left[ \beta T_{\mathrm{exit}} - \mathcal{C}(n(\cdot), k(\cdot)) \right].
\end{equation}
Assuming the KL control cost, we can derive an analytical solution. Using the fact that the terminal state $n_A(T_{\mathrm{exit}})$ is either $0$ or $N$,} the value function, i.e., the cumulant generating function of the extinction time, can be decomposed as
\ADD{
\begin{equation}\label{eq:value_function_Moran_analytical}
\begin{aligned}
	V(n, \beta) 
	&= \log \mathbb{E}_{k^0}\left[\exp(\beta T_{\mathrm{exit}})|n(0)=n\right]\\
	&= \log \mathbb{E}_{k^0}\left[\exp(\beta T_{\mathrm{exit}}) \right.\\
	&\quad\quad\quad\left.(\delta_{n_A(T_{\mathrm{exit}}),0} + \delta_{n_A(T_{\mathrm{exit}}),N}) |n(0)=n\right]\\
	&= \log \left(e^{K_0(n,\beta)} + e^{K_N(n,\beta)}\right),
\end{aligned}
\end{equation}
where 
\begin{equation}
\begin{aligned}
	K_0(n, \beta) &:=\log\mathbb{E}_{k^0}\left[\exp(\beta T_{\mathrm{exit}})\delta_{n_A(T_{\mathrm{exit}}),0} | n(0)=n\right],\\
	K_N(n,\beta) &:= \log \mathbb{E}_{k^0}\left[ \exp(\beta T_{\mathrm{exit}}) \delta_{n_A(T_{\mathrm{exit}}),N} | n(0)=n\right]
\end{aligned}
\end{equation}
} are the cumulant-generating functions of the first hitting time at $0$ and $N$, respectively. They have an explicit expression \cite{ashcroft-2015-WhenMeanNot} using the eigenvalues of submatrices of the transition rate matrix $M := (-\omega_{n,n^\prime})_{n_A,n_A^\prime\neq 0, N} \in \mathbb{R}^{(N-1)\times(N-1)}$ as follows:
\begin{equation}\label{eq:cgf_Moran_0}
\begin{aligned}
	K_0(n, \beta)
	&=\sum_{a = 1}^{n_A} \log d^0(a) + \sum_{i=1}^{N-n_A-1} \log(x_i(n_A) - \beta)\\
	&\quad- \sum_{k=1}^{N-1}\log(\lambda_k - \beta),
\end{aligned}
\end{equation}
and
\begin{equation}\label{eq:cgf_Moran_N}
\begin{aligned}
	K_N(n, \beta)
	&=\sum_{a = n_A}^{N-1} \log b^0(a) + \sum_{j=1}^{n_A-1} \log(y_j(n_A) - \beta)\\
	&\quad - \sum_{k=1}^{N-1} \log(\lambda_k - \beta),
\end{aligned}
\end{equation}
where $b^0(a):=k_1^0 h_1(a, N-a)$ and $d^0(a):=k_2^0 h_2(a, N-a)$. In the equations, $x_i(n_A) > 0$, $y_j(n_A) > 0$, and $\lambda_k > 0$ are the eigenvalues of bottom-right submatrix of size $(N-n_A-1)$, top-left submatrix of size $(n_A - 1)$, and the full matrix $M$, respectively \cite{ashcroft-2015-WhenMeanNot}.
Due to the interlacing property \cite{hershkowitz-1987-EigenvalueInterlacingCertain}, the smallest eigenvalue $\lambda_1$ of the full matrix is smaller than the smallest eigenvalues, $x_1(n)$ and $y_1(n)$, of the submatrices. 
Therefore, the value function diverges $V(n, \beta) \rightarrow +\infty$ as $\beta\rightarrow\lambda_1$. 
The expected extinction time $\mathcal{U}^\dagger(n, \beta)$ under the optimal control also diverges as $\beta \rightarrow \lambda_1$ because it is the derivative of $V(n, \beta)$ by $\beta$ (Eq.~\eqref{eq:expected_optimal_utility}).
This critical $\beta_c := \lambda_1$ is the same for all initial conditions $n$.



We numerically calculated the value function in two ways: by solving the linear algebraic Eq.~\eqref{eq:fisrt_exit_linear_HJB_equation} and by using the analytical formula in Eqs.~\eqref{eq:value_function_Moran_analytical}–\eqref{eq:cgf_Moran_N}. Two solutions exactly match, and the value function $V(n, \beta)$ diverges as $\beta$ approaches the smallest eigenvalue $\beta\rightarrow\beta_c=\lambda_1$ (Fig.~\ref{fig:Moran_first_exit}~(a)).
The cost-exit time tradeoff curves $\mathcal{E}(n, u)$ in Fig.~\ref{fig:Moran_first_exit}~(b) approach lines with slope $\beta_c$ in the long exit time regime. Thus, the cost rate per unit time $\mathcal{E}(n, u) / u$ is upper bounded by $\beta_c$ as shown in Fig.~\ref{fig:Moran_first_exit}~(c). 
The result indicates that a finite amount of control cost per time is sufficient to prevent the ultimate extinction on average. 
As $\gamma = k^0_1 / k^0_2$ increases, the extinction tends to happen earlier, resulting in the increased control cost rate $\beta_c$ for preventing extinction.

As a function of $n_A$, the value function $V(n, \beta)$ for each $\beta$ and $\gamma$ has a single peak, which we designate by $n_A^*$ (Fig.~\ref{fig:Moran_first_exit}~(d)). 
The optimal control steers the system to stay around $n_A^*$.
As $\beta$ increases, the derivative of the value function becomes steep around 
 $n_A^*$ (Fig.~\ref{fig:Moran_first_exit}~(e)) so that the controlled trajectories do not leave there even after a long time (Fig.~\ref{fig:Moran_first_exit}~(f)).

When $\beta$ is close to $\beta_c$, the width of the peak around $n_A^*$ decreases as $\gamma$ increases, and there emerges a region with a shallow slope $\overline{\nabla}_{s_1}V(n, \beta) \approx 0$. 
In this region, the optimal control $k^\dagger_r(n) = k^0_r \exp(\overline{\nabla}_{s_r}V(n, \beta))$ become close to the uncontrolled rate $k^0_r$.
The emergence of the shallow slope region indicates that the optimal control strategy switches between ON and OFF modes depending on the current state.

The OFF mode region appears if $\gamma$ is high and $N$ is moderately large (see Appendix~\ref{appendix:moran_control}),
implying that the OFF mode is attributed to the difficulty of controlling an exponentially growing population. 
If both $\gamma$ and $N$ are large, the uncontrolled system has a stronger and less stochastic driving force towards $n_A=N$ as $n_A$ increases.
Thus, the optimal control keeps the population away from $n_A=N$ by forcing it close to $0$, which is the ON mode around the peak $n_A^*$.
Once a large fluctuation drives the population to leave the peak, the additional cost of bringing it back to the peak does not compensate for its success rate, leading to the OFF mode, i.e., doing nothing, in the intermediate region. 
Finally, the control turns ON again near $n_A=N$ to hang on there. 
Such a spontaneous emergence of hierarchical control may not be identifiable within the LQG approximation, highlighting the importance of considering the unique properties of RN.

\ADD{
The result explained so far used the KL control cost problem. The existence of a finite cost rate to prevent ultimate extinction and the emergence of the mode-switching control strategy could be attributed to either the special choice of the control cost function, the problem itself (maximizing extinction time for Moran processes), or both.
To reveal the effect of the control cost function, we numerically solved the optimal control problem with two types of non-KL control cost functions.
The first control cost function is the quadratic control cost $C_{quad}$ given by Eq.~\eqref{eq:quadratic_control_cost}. While the quadratic and KL control cost functions are the same up to the second order in $k-k^0$ around $k=k^0$, they behave differently in regions far from $k^0$.
The second class of control cost functions is the weighted KL control cost given by
\begin{equation}
	C_{wKL}(n, k) = \sum_{r \in R} w_r h_r(n) \left(k_r \log \frac{k_r}{k_r^0} - k_r + k_r^0\right),
\end{equation}
where $w_r > 0$ is the weighting factor for each reaction. When the weighting factor for reaction $r$ is infinite, i.e., $w_r = +\infty$, the reaction rate coefficient $k_r$ is constrained to be $k_r^0$.
}

\ADD{
We obtained the optimal solutions by numerically solving the nonlinear HJB Eq.~\eqref{eq:HJB_first_exit_f}.
As shown in Fig.~\ref{fig:Moran_other_cost}~(a-c), the value function $V(n,\beta)$ as a function of $\beta$ and the cost-exit time tradeoff curves $\mathcal{E}(n,u)$ for the quadratic control cost are qualitatively similar to that of KL control cost. The existence of a critical cost rate, $\beta_c$, that prevents ultimate extinction can be confirmed.
}
\ADD{
Furthermore, as shown in Fig.~\ref{fig:Moran_other_cost}~(d-f), the value functions as a function of $n_A$ with the various non-KL control costs are qualitatively similar to those of the KL control cost.
Despite the quantitative difference, the derivative of the value function always exhibits the OFF mode region. Consequently, the simulated controlled trajectories show the mode-switching behavior. Therefore, the emergence of the mode-switching strategy is not due to the nature of the KL control cost, but rather an intrinsic feature of the control problem for Moran processes.
}

\subsection{Controlling epidemic outbreak}
\begin{figure*}[htp]
\centering
\includegraphics[width=18cm]{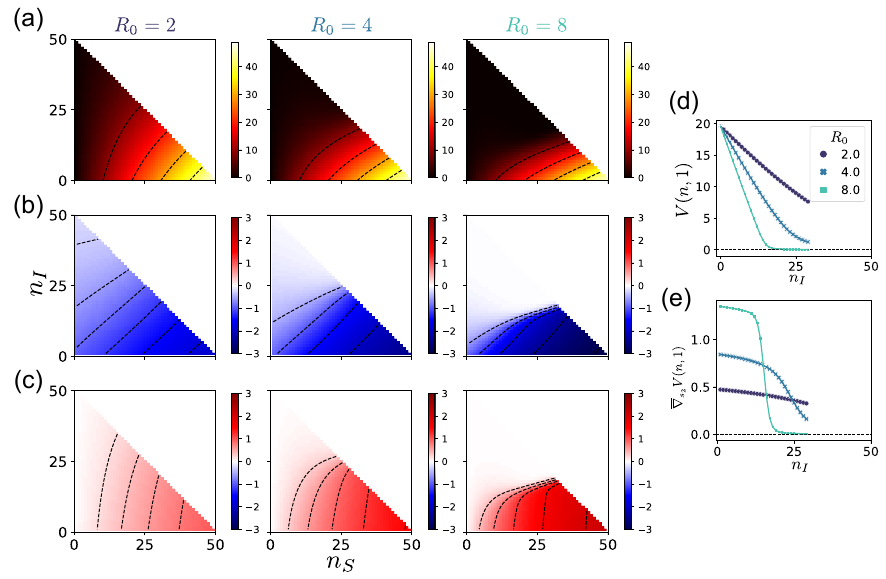}
\caption{The solution of the minimum total infection problem for stochastic SIR model for different values of $R_0\in\{2,4,8\}$. (a) Heat maps of the value function $V(n, 1)$ plotted on the $n_S$–$n_I$ plane with $\beta = 1$ for varying $R_0$. Light color represents a large value. 
(b) The directional derivative $\overline{\nabla}_{s_1}V(n, 1)$ of the value function in the infection direction $s_1$, which determines the optimally controled infection rate as $k^\dagger_1(n) = k^0_1 \exp(\overline{\nabla}_{s_1} V(n, 1))$. 
(c) The directional derivative $\overline{\nabla}_{s_2}V(n, 1)$ of the value function in the recovery direction $s_2$, which determines the optimally controled recovery rate as $k^\dagger_2(n) = k^0_2 \exp(\overline{\nabla}_{s_2} V(n, 1))$. 
(d) The value function $V(n, 1)$ on the line $n_S = 20$. (e) The derivative of the value function $\overline{\nabla}_{s_2}V(n, 1)$. The parameters are $N=50$, $\beta = 1$, $k^0_2=0.01$, $k^0_1 = R_0 k^0_2$.}
\label{fig:SIR_first_exit}
\end{figure*}

\begin{figure*}[htp]
\centering
\includegraphics[width=18cm]{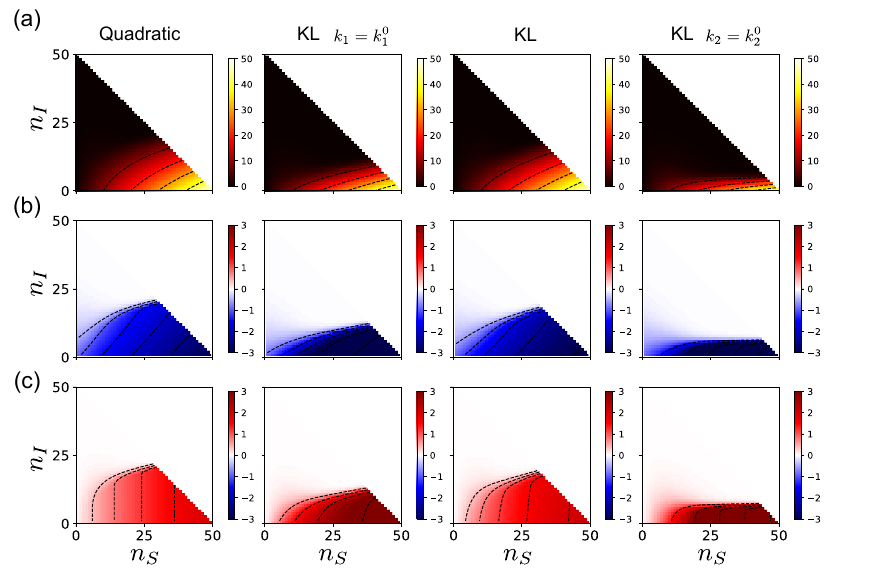}
\caption{\ADD{
Comparison of the minimum total infection solution for the stochastic SIR model with different control cost functions. 
(a) The value function $V(n, 1)$ (b,c) its derivative $\overline{\nabla}_{s_1}V(n, 1)$ and $\overline{\nabla}_{s_2}V(n, 1)$ with the quadratic $C_{quad}$, weighted KL $C_{wKL}$ ($w_1=\infty, w_2=1$), KL $C_{KL}$, and weighted KL $C_{wKL}$ ($w_1=1, w_2 = \infty$) control cost functions.
Parameters are $N=50$, $\beta = 1$, $k^0_2=0.01$, $k^0_1 = R_0 k^0_2$, $R_0 = 8$.
}
}
\label{fig:SIR_other_cost}
\end{figure*}

Lastly, we apply our framework to epidemic problems.
We use the stochastic SIR model (Fig.~\ref{fig:intro} (c)) in which the population is divided into three classes: susceptible ($S$), infected ($I$), and recovered ($R$), i.e., $X=\{ S, I, R\}$. The uncontrolled process is described by the following infection and recovery reactions:
\begin{equation}
	S + I \overset{k_1}{\longrightarrow} 2I,
	\quad
	I \overset{k_2}{\longrightarrow} R,
\end{equation}
and we assume mass-action-type kinetics, i.e., $h_1(n) = N^{-1}n_S n_I$ and $h_2(n) = n_I$.
The total population size $N := n_S(t)+n_I(t)+n_R(t)$ is a conserved quantity of the system.
This model has absorbing states $\mathcal{N}_{\mathrm{abs}} = \{ n \in \mathcal{N} | n_I = 0 \}$ and any stochastic solution $n(t)$ eventually reaches one of these states as $t\rightarrow \infty$.
When the ratio $R_0 := k_1^0 / k_2^0$ between reaction rate coefficients is high, the infection spreads rapidly in the population, and the state $n(T_{\mathrm{exit}})$ at the terminal time tends to have a small number $n_S(T_{\mathrm{exit}})$ of susceptible and a large number $n_R(T_{\mathrm{exit}})$ of recovered people. The number of recovered people at the end is equal to the total number of infections during the epidemic, which is known as the size of the epidemic~\cite{bailey-1953-TotalSizeGeneral}.

The goal of control is then to minimize the size of the epidemic or, equivalently, to maximize the number of susceptible people at the end of the epidemic.
Let us formulate it as the first exit problem where $U_{\mathrm{exit}}(n)=n_S$ and $U(n)\equiv 0$, i.e.,\ADD{
\begin{equation}
	\underset{k(\cdot)}{\text{maximize}} \quad \beta \mathbb{E}_{k(\cdot)} \left[ n_S(T_{\mathrm{exit}}) - \mathcal{C}(n(\cdot), k(\cdot)) \right].
\end{equation}}

\ADD{To begin with, we assumed the KL control cost and} numerically calculated the value function and the optimal reaction rate using Eqs.~\eqref{eq:fisrt_exit_linear_HJB_equation} and \eqref{eq:optimal_reaction_rate} for $N=50$.
The results are shown in Fig.~\ref{fig:SIR_first_exit}.
The value decreases as the number of susceptible people decreases and as the number of infected people increases (Fig.~\ref{fig:SIR_first_exit}~(a)). 
The optimal control of infection and recovery rates plotted in Fig.~\ref{fig:SIR_first_exit} (b) and (c) indicates that strong control to reduce the infection rate and to boost the recovery rate is encouraged when the number of susceptibles is large and the number of infected is small.
However, if many people are already infected, strong control is no longer encouraged. Instead, almost no control over the infection and recovery rates becomes optimal, as indicated by the white regions in Fig.~\ref{fig:SIR_first_exit} (b) and (c).
This means that almost the entire population will eventually become infected in this situation, no matter how optimally the rates are controlled, and that further investment in the control cost is not worth the potential gains.

In particular, we can observe a sharp transition between strong control and no control when $R_0$ is high.
Figures~\ref{fig:SIR_first_exit}~(d) and (e) show the value function $V(n, 1)$ and its derivative $\overline{\nabla}_{s_2} V(n, 1)$ on the line $n_S=20$, which determines the optimal recovery rate $k^\dagger_2(n) = k^0_2 \exp(\overline{\nabla}_{s_2} V(n, 1))$. 
The value functions are approximately piecewise-linear functions $V(n, 1)\approx \max\{ 0, a_0 n_S + a_1 n_I + a_2 \}$, so the derivatives and the optimal rates are approximately piecewise-constant. This transition leads to the mode switching of optimal control, which is similar to the case of the problem of maintaining diversity.

\ADD{
As in the previous section, we then calculated the optimal solutions with the quadratic and weighted KL control cost functions with high $R_0=8$. Figure~\ref{fig:SIR_other_cost} demonstrates the mode-switching of the optimal strategies with the quadratic and weighted KL control cost functions, as is observed with the KL control cost.
Therefore, the mode switching of the optimal control is also an intrinsic feature of the minimum infection problems rather than the artifact of the KL control cost.
}

\section{Discussion}
In this paper, we propose a new framework for formulating optimal control problems of stochastic reaction networks. 
We show that the \ADD{$f$-divergence  possesses desirable properties as a control cost for reaction networks. 
In particular, when the control cost is given by the Kullback–Leibler (KL) divergence}, the associated Hamilton–Jacobi–Bellman (HJB) equation can be linearized via the Cole–Hopf transformation, yielding both analytical insight and computational efficiency.
We demonstrate the effectiveness of our framework through applications to three classes of biological control problems involving absorbing states.

There are several potential directions worth investigating.
First, our framework can be extended to risk-sensitive control problems \cite{jacobson-1973-OptimalStochasticLinear,whittle-1990-RiskSensitiveOptimal}, where not only the expectation of performance but also its variance and higher-order moments are considered. For instance, as the concentration of intracellular molecules inevitably fluctuates, it is vital to suppress fluctuations and variability when robust homeostasis is required \cite{raser-2005-NoiseGeneExpression}. As studied for diffusion processes \cite{broek-2012-RiskSensitivePath}, risk sensitivity can be incorporated into the current framework.

Second, more realistic models of reaction networks may involve many species or large population sizes, as seen in complex ecological systems and stage-structured epidemic models \cite{xiao-2020-EcologicalFrameworkUnderstand,skwara-2023-StatisticallyLearningFunctional,brauer-2012-MathematicalModelsPopulationa,dsouza-2023-ControllingComplexNetworks}.
The optimal control of such large-scale systems leads to high-dimensional equations. Although the HJB equation becomes linear under KL control, solving it remains computationally demanding.
The development of fast and scalable numerical algorithms is therefore essential.
Sampling-based techniques \cite{kappen-2005-PathIntegralsSymmetry,theodorou-2010-ReinforcementLearningMotor} may offer efficient solutions in such high-dimensional settings.

\ADD{
Third, the computational advantages afforded by the Cole–Hopf transformation do not extend to general $f$-divergence control costs beyond the KL control cost.}
This limitation is analogous to the condition in diffusion control problems, where efficient solutions require the control cost weight to be inversely proportional to the noise intensity~\cite{kappen-2005-PathIntegralsSymmetry,todorov-2009-EfficientComputationOptimal}.
For stochastic reaction networks, the noise strength is linked to the propensity function $h_r(n)$, which should be used as the weight in the control cost. 
\ADD{
As demonstrated in our examples involving the Moran and SIR models with non-KL control costs, solving the nonlinear HJB equations numerically is tractable for small-scale problems.}
For large-scale problems, iterative methods with local approximation, as in \cite{todorov-2005-GeneralizedIterativeLQG,satoh-2017-IterativeMethodNonlinear}, might provide a way to overcome this limitation.

\ADD{
Finally, the assumption of full observability of the current state does not hold in realistic scenarios.
When the feedback controller does not have direct access to the true state, the optimal control problem must be reformulated to include state estimation via a filtering equation, followed by control based on the resulting belief state.
Not only control but also filtering for the stochastic reaction networks is challenging due to their nonlinearity and non-Gaussian randomness~\cite{rathinam-2021-StateParameterEstimation,fang-2022-StochasticFilteringMultiscale,hammouda-2025-FilteredMarkovianProjection}.
Meanwhile, a study~\cite{nakamura-2022-OptimalSensingControl} investigated a partially observable optimal control problem for Markov jump processes with Gaussian observation noise, demonstrating that the KL control cost can be useful in deriving the optimal solution even in the partially observable setting. 
Developing practical algorithms for filtering and control in general stochastic reaction networks remains a promising direction for future work.
}

\begin{acknowledgments}
We thank Simon Schnyder, Louis-Pierre Chaintron, and Kenji Kashima for their helpful discussions.
This research is supported by JST CREST JPMJCR2011 and JPMJCR1927, and JSPS KAKENHI Grant Numbers 24KJ0090, 24H02148, 25H01365.
\end{acknowledgments}

\onecolumngrid
\appendix

\section{Details about optimal control problem}\label{appendix:derivation_optimal_control}
In this section, we present the general formulation of the optimal control problems in detail and then show it when the Kullback–Leibler (KL) control cost is employed.

Firstly, we summarize the functions that specify the problem. The terminal utility function $U_T(n)$ and the instantaneous utility function $U_t(n)$ are scalar functions of the state $n\in \mathcal{N}$. The instantaneous control cost function $C(n, k)$ is a scalar function of the state $n\in\mathcal{N}$ and the reaction rate coefficient $k\in\mathbb{R}_{\geq 0}^{|R|}$.
Moreover, the instantaneous control cost function $C(n, k)$ is assumed to be continuous, strictly convex, and 1-coercive with respect to $k$ for each $n\in \mathcal{N}$, i.e., 
\begin{equation}
	C(n, a k + (1-a) k^\prime) < a C(n, k) + (1-a) C(n, k^\prime)
\end{equation}
for any $a \in (0,1)$ and $k, k^\prime \in \mathbb{R}_{\geq 0}^{|R|}$, and 
\begin{equation}
	\lim_{a \rightarrow +\infty}\frac{C(n, a k)}{a} = +\infty
\end{equation}
for any $k\in \mathbb{R}_{\geq 0}^{|R|}\backslash\{0\}$.
The control cost function with these assumptions is general enough in practical situations, and they guarantee that the optimal reaction rate coefficients uniquely exist. 
Furthermore, since adding a constant does not affect the optimal solution, we assume that the instantaneous control cost function is nonnegative, $C(n, k) \geq 0$. Since $C(n, k)$ defined above always has a unique minimizer $k^0(n)$ for each $n\in \mathcal{N}$, we assume that $C(n, k^0(n)) = 0$.

From these three functions, the utility function $\mathcal{U}(n(\cdot))$, a scalar function of the state trajectory $(n(t))_{0\leq t \leq T}$ from time $0$ to $T$ is defined as
\begin{equation}
	\mathcal{U}(n(\cdot)) := U_T(n(T)) + \int_0^T U_t(n(t)) dt.
\end{equation}
Note that we denote the trajectory $(n(t))_{0\leq t \leq T}$ as $n(\cdot)$ when the time interval $[0, T]$ is evident from the context.
The control cost function $\mathcal{C}(n(\cdot), k(\cdot))$ is defined similarly as
\begin{equation}
	\mathcal{C}(n(\cdot), k(\cdot)) := \int_0^T C(n(t), k(t))dt.
\end{equation}

First, we consider a constrained optimization problem
\begin{equation}
\begin{aligned}\label{eq:conjugate_value_function_0_appendix}
	\mathcal{E}(n,u) :=~&\underset{k(\cdot)}{\min} ~~ \mathbb{E}_{k(\cdot)}\left[ \mathcal{C}(n(\cdot), k(\cdot)) | n(0)=n\right]\\
	&\text{s.\,t.}~~ \mathbb{E}_{k(\cdot)} \left[ \mathcal{U}(n(\cdot)) | n(0)=n\right] \geq u,
\end{aligned}
\end{equation}
where the expectation $\mathbb{E}_{k(\cdot)}$ is taken over the trajectories $n(\cdot)$ generated by the stochastic RN with $k(t)$ as the reaction rate coefficients at time $t$. The problem is to find the optimal trajectory of reaction rate coefficient $k(\cdot)$ that minimizes the expected control cost such that the expected utility is equal to the given value $u\in\mathbb{R}$.

Introducing the Lagrange multiplier $\beta \geq 0$, the above problem is equivalent to the following unconstrained optimization problem
\begin{equation}\label{eq:value_function_0_appendix}
	V(n, \beta) := \max_{k(\cdot)}  \mathbb{E}_{k(\cdot)} \left[ \beta \mathcal{U}(n(\cdot)) - \mathcal{C}(n(\cdot), k(\cdot)) | n(0)=n\right].
\end{equation}
Once we can solve this problem, the solution to the original problem can be obtained. Thus, we focus on solving the unconstrained problem, Eq.~\eqref{eq:value_function_0_appendix}. 

We can derive the Hamilton–Jacobi–Bellman (HJB) equation to find the optimal reaction rates $(k^\dagger(t))_{0 \leq t < T}$ and the maximum value $V(n, \beta)$ by following the standard procedure in the control theory, cf., \cite{boel-1977-OptimalControlJump}. 
Let the value function $V_t(n,\beta)$ for $0\leq t \leq T$, $n\in \mathcal{N}$ and $\beta > 0$ be defined as the above objective function from $t$ to $T$ conditioned on the state $n(t)$ at time $t$:
\begin{equation}\label{eq:value_function_appendix}
	V_t(n,\beta) := \max_{k(\cdot)} \mathbb{E}_{k(\cdot)}\left[
	\beta \mathcal{U}_t(n(\cdot)) - \mathcal{C}_t(n(\cdot), k(\cdot))
	| n(t) = n \right],
\end{equation}
where $\mathcal{U}_t(n(\cdot))$ and $\mathcal{C}_t(n(\cdot), k(\cdot))$ are the utility and control cost functions from time $t$ to $T$, i.e., $\mathcal{U}_t(n(\cdot)) := U_T(n(T)) + \int_t^T U_\tau(n(\tau))d\tau$ and $\mathcal{C}_t(n(\cdot), k(\cdot)) := \int_t^T C_\tau(n(\tau), k(\tau))d\tau$.
By applying the dynamic programming principle, the HJB equation can be derived as, for $t \in (0,T)$,
\begin{equation}\label{eq:HJB_appendix}
	-\frac{\partial}{\partial t}V_t(n,\beta) = \beta U_t(n) + \max_{k\in \mathbb{R}_{\geq 0}^{|R|}} \left[ \mathcal{L}_k V_t(n,\beta) - C(n, k) \right],
\end{equation}
with the terminal condition $V_T(n,\beta) = \beta U_T(n)$. $\mathcal{L}_k$ is the backward operator which is given by, for any scalar function $f(n)$,
\begin{equation}
	\mathcal{L}_k f(n) = \sum_{r\in R} k_r h_r(n) \overline{\nabla}_{s_r} f(n),
\end{equation}
where $\overline{\nabla}_{s_r} f(n) := f(n + s_r) - f(n)$ is the discrete directional derivative.
Due to the strict convexity and coercivity of the instantaneous control cost function, Eq.~\eqref{eq:HJB_appendix} can be rewritten as
\begin{equation}\label{eq:HJB_conjugate_appendix}
	-\frac{\partial}{\partial t}V_t(n,\beta) = \beta U_t(n) + C^*(n, h(n)\overline{\nabla}_s V_t(n, \beta)),
\end{equation}
where $C^*(n, \phi) := \max_k [\sum_{r\in R} k_r \phi_r - C(n, k)]$ is the Legendre transform of $C(n,k)$ with respect to $k$.
The optimal reaction rate coefficient $k^\dagger(t, n, \beta)$ is given by the following time-dependent state-feedback controller
\begin{equation}\label{eq:optimal_control_conjugate_appendix}
	k^\dagger(t, n, \beta) = \partial_\phi C^*(n, h(n)\overline{\nabla}_s V_t(n, \beta)),
\end{equation}
where $\partial_{\phi} C^*(n, \phi) \in \mathbb{R}^{|R|}$ is the derivative of $C^*(n, \phi)$ with respect to $\phi \in \mathbb{R}^{|R|}$.

Plugging the optimal reaction rate coefficient into Eq.~\eqref{eq:value_function_appendix} and taking the derivative with respect to $\beta$, one obtains the expected utility under the optimal control $k^\dagger$ as
\begin{equation}\label{eq:optimal_utility_appendix}
\begin{aligned}
	\mathcal{U}^\dagger_t(n, \beta) 
	&:= \mathbb{E}_{k^\dagger(\cdot,\beta)}\left[\mathcal{U}_t(n(\cdot)) | n(t)=n\right]\\
	&= \partial_\beta V_t(n, \beta).
\end{aligned}
\end{equation}
The optimal expected utility $\mathcal{U}^\dagger_t(n,\beta)$ is an increasing function of $\beta$ because the value function is convex with respect to $\beta$. The convexity can be confirmed as follows: for any $0\leq a \leq 1$ and $\beta_1, \beta_2 \in \mathbb{R}$,
\begin{equation}
\begin{aligned}
	V_t&(n, a\beta_1 + (1-a)\beta_2) \\
	&= \max_{k(\cdot)}\mathbb{E}_{k(\cdot)}\left[(a\beta_1 + (1-a)\beta_2)\mathcal{U}_t(n(\cdot)) - \mathcal{C}_t(n(\cdot), k(\cdot))\right]\\
	&= \max_{k(\cdot)} \left[a\mathbb{E}_{k(\cdot)}\left[\beta_1 \mathcal{U}_t(n(\cdot)) - \mathcal{C}_t(n(\cdot), k(\cdot))\right] +
	(1-a)\mathbb{E}_{k(\cdot)}\left[\beta_2 \mathcal{U}_t(n(\cdot)) - \mathcal{C}_t(n(\cdot), k(\cdot))\right]\right]\\
	&\leq a\max_{k(\cdot)} \mathbb{E}_{k(\cdot)}\left[\beta_1 \mathcal{U}_t(n(\cdot)) - \mathcal{C}_t(n(\cdot), k(\cdot))\right] + 
	(1-a)\max_{k(\cdot)}\mathbb{E}_{k(\cdot)}\left[\beta_2 \mathcal{U}_t(n(\cdot)) - \mathcal{C}_t(n(\cdot), k(\cdot))\right]\\
	&= a V_t(n,\beta_1) + (1-a) V_t(n, \beta_2).
\end{aligned}
\end{equation}

On the other hand, the expected cost under the optimal control is given by
\begin{equation}
\begin{aligned}
	\mathcal{C}^\dagger_t(n, \beta) 
	&:= \mathbb{E}_{k^\dagger(\cdot, \beta)}\left[\mathcal{C}_t(n(\cdot), k^\dagger(\cdot,\beta)) | n(t)=n\right]\\
	&= \beta \mathcal{U}^\dagger_t(n,\beta) - V_t(n,\beta).
\end{aligned}
\end{equation}
These formulations imply that the optimal expected cost $\mathcal{C}^\dagger_t(n,\beta)$ and the optimal expected utility $\mathcal{U}^\dagger_t(n,\beta)$ are tightly related by the Legendre transform of the value function $V_t(n,\beta)$ with respect to $\beta$. For $u \in \mathbb{R}$, we define {\it conjugate value function} as
\begin{equation}
	\mathcal{E}_t(n, u) := \sup_{\beta \geq 0}\left[ \beta u - V_t(n, \beta) \right].
\end{equation}
Then,
\begin{equation}\label{eq:lagrange_sup_inf}
\begin{aligned}
	\mathcal{E}_t(n, u) 
	&= \sup_{\beta \geq 0}\inf_{k(\cdot)}\left[ \beta u - \mathbb{E}_{k(\cdot)}\left[\beta \mathcal{U}_t(n(\cdot)) - \mathcal{C}_t(n(\cdot),k(\cdot)) | n(t)=n\right] \right]\\
	&= \sup_{\beta \geq 0}\inf_{k(\cdot)} L(\beta, k(\cdot)),
\end{aligned}
\end{equation}
where the Lagrange function $L(\beta, k(\cdot))$ is defined as
\begin{equation}
\begin{aligned}
	L(\beta, k(\cdot)) := \beta\left( u - \mathbb{E}_{k(\cdot)}\left[  \mathcal{U}_t(n(\cdot))|n(t)=n\right]\right)
	+ \mathbb{E}_{k(\cdot)}\left[\mathcal{C}_t(n(\cdot),k(\cdot)) |n(t)=n\right].
\end{aligned}
\end{equation}
If the target expected utility $u$ is achievable by $\beta=\beta^*$, i.e., there exists $\beta^* \geq 0$ such that
\begin{equation}\label{eq:existence_u_beta}
	\mathbb{E}_{k^\dagger(\cdot,\beta^*)}\left[\mathcal{U}_t(n(\cdot)) | n(t)=n\right] = u
\end{equation}
holds, then $(\beta^*, k^\dagger(\cdot, \beta^*))$ is the saddle point of the function $L$, i.e, for any $\beta$ and $k(\cdot)$,
\begin{equation}\label{eq:saddle_point_condition}
	L(\beta, k^\dagger(\cdot, \beta^*))
	\leq L(\beta^*, k^\dagger(\cdot, \beta^*))
	\leq L(\beta^*, k(\cdot)).
\end{equation}
The second inequality holds because of the definition of the optimal control $k^\dagger$ and the first inequality holds because
\begin{equation}
\begin{aligned}
	L&(\beta, k^\dagger(\cdot, \beta^*)) - L(\beta^*, k^\dagger(\cdot, \beta^*))\\
	&= (\beta - \beta^*)\underbrace{\left( u - \mathbb{E}_{k^\dagger(\cdot,\beta^*)}\left[  \mathcal{U}_t(n(\cdot))|n(t)=n\right]\right)}_{= 0 (\because \text{Eq.~\eqref{eq:existence_u_beta}})}\\
	&\leq 0.
\end{aligned}
\end{equation}
The saddle point property in Eq.~\eqref{eq:saddle_point_condition} justifies exchanging the order of maximization and minimization \cite{boyd-2004-ConvexOptimization}, and Eq.~\eqref{eq:lagrange_sup_inf} becomes
\begin{equation}
\begin{aligned}
	\mathcal{E}_t(n, u) 
	&= \sup_{\beta \geq 0}\inf_{k(\cdot)} L(\beta, k(\cdot))\\
	&= \inf_{k(\cdot)}\sup_{\beta \geq 0}L(\beta, k(\cdot))\\
	&= \inf_{k(\cdot)}\left[\mathbb{E}_{k(\cdot)}\left[\mathcal{C}_t(n(\cdot),k(\cdot)) |n(t)=n\right]
	+ \sup_{\beta \geq 0}\beta\left( u - \mathbb{E}_{k(\cdot)}\left[  \mathcal{U}_t(n(\cdot))|n(t)=n\right]\right) \right]\\
	&= \inf_{k(\cdot)}\left[\mathbb{E}_{k(\cdot)}\left[\mathcal{C}_t(n(\cdot),k(\cdot)) |n(t)=n\right] + \iota_{K_u}(k(\cdot))\right]\\
	&= \underset{k(\cdot)\in K_u}{\inf}  \mathbb{E}_{k(\cdot)}\left[ \mathcal{C}_t(n(\cdot), k(\cdot)) | n(t)=n\right]
\end{aligned}
\end{equation}
where the indicator function $\iota_A(a)$ of a set $A$ is defined as $\iota_{A}(a)=0$ when $a\in A$ and $\iota_{A}(a)=+\infty$ otherwise, and $K_u$ is the set of reaction rate coefficients $k(\cdot)$ which satisfies $\mathbb{E}_{k(\cdot)}\left[  \mathcal{U}_t(n(\cdot))|n(t)=n\right] \geq u$.

Therefore, the conjugate value function gives the minimum expected cost under the constraint on the expected utility, i.e., 
\begin{equation}
\begin{aligned}\label{eq:conjugate_value_function_appendix}
	\mathcal{E}_t(n,u) =~&\underset{k(\cdot)}{\text{min}} ~~ \mathbb{E}_{k(\cdot)}\left[ \mathcal{C}_t(n(\cdot), k(\cdot)) | n(t)=n\right]\\
	&\text{s.\,t.}~~ \mathbb{E}_{k(\cdot)} \left[ \mathcal{U}_t(n(\cdot)) | n(t)=n\right] \geq u.
\end{aligned}
\end{equation}
The solution to the original constrained optimization problem in Eq.~\eqref{eq:conjugate_value_function_0_appendix} can be obtained as $\mathcal{E}(n, u) = \mathcal{E}_0(n, u)$.
In other words, once we have the value function for all $\beta$, we can calculate the relation between the optimal expected utility and the optimal expected cost.

The conjugate value function gives an intuitive picture of the relation between performance and required cost, which often yields a tradeoff relation.
For instance, the conjugate value function is minimized in the uncontrolled situation $\beta = 0$ and maximized in the strongly controlled situation $\beta \rightarrow +\infty$.
This is because the Legendre duality between the value function $V_t(n,\beta)$ and the conjugate value function $\mathcal{E}_t(n,u)$ induces the correspondence between the weight $\beta$ and the constraint $u$ which is given by the following equation
\begin{equation}
	u = \partial_\beta V_t(n, \beta) = \mathcal{U}_t^\dagger(n, \beta)
\end{equation}
and its inverse
\begin{equation}
	\beta = \partial_u \mathcal{E}_t(n, u)
\end{equation}
if the conjugate value function is differentiable.
The minimizer $u_0$ of the conjugate value function has a vanishing derivative $\partial_u \mathcal{E}_t(n, u_0)=0$, which is equivalent to $u_0 = \partial_\beta V_t(n,0) = \mathcal{U}_t^\dagger(n, 0)$ meaning that the uncontrolled case $\beta=0$ corresponds to the minimizer of the conjugate value function. 

In some cases, there exists a maximum $\beta_c$ such that the value function $V(n,\beta)$ diverges to $+\infty$ as $\beta \rightarrow \beta_c$. In this case, the conjugate value function $\mathcal{E}(n, u)$ has an asymptotic slope $\beta_c$ as $u \rightarrow \infty$.
On the other hand, the value function can have a finite asymptotic slope $u_c$ as $\beta \rightarrow \infty$. This is when the conjugate value function diverges to $+\infty$ at $u=u_c$.

Finally, let us apply the formulations above when the instantaneous control cost function $C(n, k)$ is of KL type as defined in Eq.~\eqref{eq:KL_cost_function}. It is a continuous, strictly convex, 1-coercive, and nonnegative function minimized at $k=k^0$. The Legendre transform of the KL control cost function $C_{KL}(n, k)$ with respect to $k$ is given by
\begin{equation}
	C_{KL}^*(n, \phi) = \sum_{r\in R} k^0_r h_r(n) \left(\exp\left(\frac{\phi_r}{h_r(n)}\right) - 1\right),
\end{equation}
where the summand is understood to be zero when $h_r(n)=0$.
Thus, the HJB equation in Eq.~\eqref{eq:HJB_conjugate_appendix} becomes
\begin{equation}\label{eq:HJB_KL_appendix}
\begin{aligned}
	-\frac{\partial}{\partial t}V_t(n,\beta) &= \beta U_t(n) \\
	&\quad+ \sum_{r\in R} k^0_r h_r(n) \left(\exp\left(\overline{\nabla}_{s_r} V_t(n, \beta)\right) - 1\right),
\end{aligned}
\end{equation}
which is presented in Eq.~\eqref{eq:HJB_for_V} in the main text.
The optimal reaction rate coefficient in Eq.~\eqref{eq:optimal_control_conjugate_appendix} becomes
\begin{equation}
	k^\dagger_r(t, n, \beta) = k_r^0 \exp(\overline{\nabla}_{s_r}V_t(n, \beta))
\end{equation}
because $\partial_{\phi_r} C^*_{KL}(n, \phi) = k_r^0 \exp(\phi_r / h_r(n))$.

Furthermore, as we discussed in the main text, we can obtain the probabilistic representation of the value function
\begin{equation}
	V_t(n, \beta) = \log\mathbb{E}_{k^0}\left[\exp(\beta \mathcal{U}_t(n(\cdot))) | n(t)=n\right].
\end{equation}
This is the cumulant generating function of $\mathcal{U}_t(n(\cdot))$. In the context of large deviation theory, the conjugate value function $\mathcal{E}_t(n, u)$ is related to the rate function.

\section{\ADD{Derivation of the $f$-divergence control cost}}
\label{appendix:f_divergence}
\ADD{
In this section, we briefly prove that the $f$-divergence is the only class of divergences with decomposability, positive homogeneity, and subadditivity. It has been shown in \cite{Csiszar1991,amari-2009-ADivergenceUniqueBelonging}, which is reproduced here with our notation.
}

\ADD{
From the decomposability, the divergence $D[j \| j^0]$ is written as
\begin{equation}
	D[j\| j^0] = \sum_{r\in R} D_r[j_r, j^0_r].
\end{equation}
}

\ADD{
Due to positive homogeneity, when $j_r^0 > 0$,
\begin{equation}\label{eq:proof_positive_homogeneity}
\begin{aligned}
	D_r[j_r, j^0_r] &= j^0_r D_r\left[\frac{j_r}{j^0_r}, 1\right]\\
	&=: j^0_r f_r\left(\frac{j_r}{j^0_r}\right).
\end{aligned}
\end{equation}
The function $f_r(x):=D_r[x,1]$ satisfies $f_r(1)=0$ and $\partial_x f_r(1) = 0$ due to the properties of the divergence. Equation~\eqref{eq:proof_positive_homogeneity} also holds when $j_r^0=0$ since we interpret the right-hand-side being $0$.
}

\ADD{
Finally, subadditivity implies, for $a, b \in [0,1]$,
\begin{equation}
	D_{r}[j_r, j^0_r] \leq D_{r}[a j_r, b j^0_r] + D_{r}[(1-a)j_r, (1-b)j^0_r].
\end{equation}
Substituting Eq.~\eqref{eq:proof_positive_homogeneity}, we get
\begin{equation}
	j^0_r f_r\left(\frac{j_r}{j^0_r}\right) \leq b j^0_r f_r\left(\frac{a j_r}{b j^0_r}\right) + (1-b)j^0_r f_r\left(\frac{(1-a)j_r}{(1-b)j^0_r}\right).
\end{equation}
Setting $x:=\frac{a j_r}{b j^0_r}$, and $y := \frac{(1-a) j_r}{(1-b)j^0_r}$, we obtain
\begin{equation}
	f(bx + (1-b)y) \leq b f(x) + (1-b)f(y).
\end{equation}
Since this inequality holds for any $x, y \geq 0$ and $b\in[0,1]$, $f$ is a convex function.
}

\section{Path measure perspective and the KL control cost}
\label{appendix:KL_path_measure}

In this section, we introduce the concept of path probability measures, from which the KL control cost can be derived naturally.

Let us denote probability measure on the space of reaction count process $(\xi(t))_{0\leq t \leq T}$ by $\mathbb{P}(\xi(\cdot))$. When we consider the reaction count process generated with a reaction rate coefficient function $(k(t))_{0 \leq t \leq T}$, we denote the probability measure by $\mathbb{P}_{k(\cdot)}(\xi(\cdot))$. The KL divergence between two probability measures $\mathbb{P}$ and $\mathbb{P}^0$ is defined as
\begin{equation}
\begin{aligned}
	\mathcal{D}_{KL} \left[ \mathbb{P} \| \mathbb{P}^0 \right]
	&:= \mathbb{E}_{\mathbb{P}}\left[ \log  \frac{d\mathbb{P}}{d\mathbb{P}^0}(\xi(\cdot))  \right]
\end{aligned}
\end{equation}
where $\frac{d\mathbb{P}}{d\mathbb{P}^0}(\xi(\cdot))$ is the Radon–Nikodým derivative and $\mathbb{E}_{\mathbb{P}}$ is the expectation under the probability measure $\mathbb{P}$.

When the probability measures $\mathbb{P}$ and $\mathbb{P}^0$ are given by a reaction rate coefficient function $k(\cdot)$ and the uncontrolled reaction rate coefficient $k^0(\cdot)=(k^0)_{0\leq t \leq T}$, respectively, then we can show that the KL divergence between them is equal to the expected KL control cost given by Eq.~\eqref{eq:KL_cost_function} integrated from time $0$ to $T$, i.e.,
\begin{equation}\label{eq:KL_divergence_equal_cost}
\begin{aligned}
	\mathcal{D}_{KL} \left[ \mathbb{P}_{k(\cdot)} \| \mathbb{P}_{k^0} \right]
	&= \mathbb{E}_{k(\cdot)}\left[ \mathcal{C}(n(\cdot), k(\cdot))\right]\\
	&= \mathbb{E}_{k(\cdot)}\left[ \int_0^T C(n(t), k(t))dt \right].
\end{aligned}
\end{equation}
See Appendix~\ref{sec:KL_reaction} for derivation.
Note that we use the probability measure on the space of the reaction count process $\xi(\cdot)$ rather than the number process $n(\cdot)$. 
The probability measure on the space of the number process has less information than that on the space of the reaction count process because the number process can be calculated by the reaction count process as in Eq.~\eqref{eq:reaction_count_to_numbers}, not vice versa.
The KL divergence for measures on number process space does not have a simple expression as in Eqs~\eqref{eq:KL_divergence_equal_cost}, and is not compatible with the linear HJB equation.

Now let us consider the following optimization problem with respect to the path probability measure $\mathbb{P}$ on the space of reaction counts $\xi(\cdot)$:
\begin{equation}\label{eq:optimization_prob_measure}
	\mathcal{V}(n_0, \beta) := \max_{\mathbb{P}} ~\beta \mathbb{E}_{\mathbb{P}} \left[ \mathcal{U}(n(\cdot)) \right] - \mathcal{D}_{KL} \left[ \mathbb{P} \| \mathbb{P}_{k^0} \right],
\end{equation}
where the initial condition is $n(0)=n_0$ almost surely.
This optimal measure problem Eq.~\eqref{eq:optimization_prob_measure} is equivalent to the optimal control problem Eq.~\eqref{eq:value_function_0_appendix} with KL control cost if the maximization with respect to $\mathbb{P}$ is restricted to the class of probability measures $\mathbb{P}_{k(\cdot)}$ generated by some reaction rate coefficient $k(\cdot)$. Thus, in general, the maximum $\mathcal{V}(n_0, \beta)$ is greater than the maximum $V(n_0, \beta)$, i.e., $\mathcal{V}(n_0, \beta) \geq V(n_0, \beta)$.
Nonetheless, we can prove that the equality 
\begin{equation}\label{eq:value_for_optimal_measure}
	\mathcal{V}(n_0, \beta) = V(n_0, \beta) = \log\mathbb{E}_{k^0}\left[ \exp(\beta\mathcal{U}(n(\cdot)))\right]
\end{equation}
holds in general and that the optimal probability measure $\mathbb{P}^\dagger$ of Eq.~\eqref{eq:optimization_prob_measure} coincides with the probability measure $\mathbb{P}_{k^\dagger(\cdot)}$ with the optimal reaction rate coefficient $k^\dagger(\cdot)$, i.e., $\mathbb{P}^\dagger = \mathbb{P}_{k^\dagger(\cdot)}$, which is given by
\begin{equation}\label{eq:optimal_measure}
	d\mathbb{P}^\dagger(\xi(\cdot)) = \frac{1}{\mathcal{Z}} \exp(\beta \mathcal{U}(n(\cdot))) d\mathbb{P}_{k^0}(\xi(\cdot)),
\end{equation}
where $\mathcal{Z} = \log \mathcal{V}(n_0, \beta)$ is the normalizing constant.
See Appendix~\ref{sec:proof_optimal_measure} for proof.
The equivalence to the optimal probability measure problem is the reason why the optimal control problem with KL control cost can be solved efficiently, especially in a time-forward manner without resorting to the dynamic programming principle \cite{theodorou-2012-RelativeEntropyFree}. The equivalence originates from the special property of the logarithm in KL divergence and does not hold for general divergences.

\subsection{KL divergence between two stochastic reaction networks}\label{sec:KL_reaction}
Here we prove Eq.~\eqref{eq:KL_divergence_equal_cost} by calculating the Radon–Nikodým derivative and KL divergence between two path measures $\mathbb{P}_{k(\cdot)}(\xi(\cdot))$ and $\mathbb{P}_{k^0}(\xi(\cdot))$.


Let us consider the filtered probability space $(\Omega, \mathcal{F}, P, \{\mathcal{F}_t\}_{t\in[0, T]})$.
Let
\begin{equation}
	\tilde{\xi}_r(t) := \xi_r(t) - \int_0^t k_r(\tau)h_r(n_{\tau}) d\tau.
\end{equation}
If $\mathbb{E}_{k(\cdot)}[\xi_r(t)]$ is finite for all $r$ and $t$, $\tilde{\xi}_t(r)$ is a martingale with respect to $\{\mathcal{F}_t\}$ \cite{anderson-2011-ContinuousTimeMarkov}, which means that for any $t, \Delta t \geq 0$
\begin{equation}
	\mathbb{E}_{k(\cdot)}\left[\tilde{\xi}_r(t+\Delta t)|\mathcal{F}_t\right] = \tilde{\xi}_r(t).
\end{equation}
In the differential form, we have
\begin{equation}\label{eq:differential_martingale_reaction}
	d\xi_r(t) = d\tilde{\xi}_r(t) + k_r(t)h_r(n_{t})dt,
\end{equation}
where $d\xi_r(t) = \sum_{i=1}^\infty \delta(\tau_i - t)$ if we write the time of the $i$-th jump as $\tau_i$ and $\delta(x)$ is the Dirac delta function.
Moreover, for any bounded function $f(n)$, the integral
\begin{equation}
	\int_0^T f(\xi(t^{-}))d\tilde{\xi}_r(t)
\end{equation}
is also a martingale, where $\xi(t^{-})$ is the reaction count just before the possible jump at time $t$.

The Radon–Nikodým derivative restricted to $\mathcal{F}_T$ is given by
\begin{equation}\label{eq:radon-nikodym}
	\frac{d\mathbb{P}_{k(\cdot)}}{d\mathbb{P}_{k^0}}(\xi(\cdot))
	= \exp \left(
	-\int_0^T \sum_{r \in R} \left(k_r(t) - k^0_r \right) h_r(n(t)) dt
	+\int_0^T \sum_{r \in R} \log \frac{k_r(t) h_r(n(t^-))}{k^0_r h_r(n(t^-))} d\xi_r(t)
	\right).
\end{equation}
The derivation of the Radon–Nikodým derivative for Markov jump processes can be found in Appendix~1.~Prop.~2.6 of \cite{kipnis-1999-ScalingLimitsInteracting}.
Substituting the Eq.~\eqref{eq:differential_martingale_reaction}, the log of the Radon–Nikodým derivative can be written with the martingale $\tilde{\xi}_r(t)$ as follows:
\begin{equation}
\begin{aligned}
	\log \frac{d\mathbb{P}_{k(\cdot)}}{d\mathbb{P}_{k^0}}
	&= -\int_0^T \sum_{r \in R} \left(k_r(t) - k^0_r \right) h_r(n(t)) dt
	+\int_0^T \sum_{r \in R} \log \frac{k_r(t)}{k^0_r} d\tilde{\xi}_r(t)\\
	&\quad +\int_0^T \sum_{r \in R} k_r(t)h_r(n(t)) \log \frac{k_r(t)}{k^0_r} dt.
\end{aligned}
\end{equation}
Thus, taking the expectation, the integral term of the martingale vanishes and get
\begin{equation}
\begin{aligned}
	\mathcal{D}_{KL} \left[ \mathbb{P}_{k(\cdot)} \| \mathbb{P}_{k^0} \right]
	&= \mathbb{E}_{k(\cdot)} \left[ -\int_0^T \sum_{r \in R} \left(k_r(t) - k^0_r \right) h_r(n(t)) dt
	 +\int_0^T \sum_{r \in R} k_r(t)h_r(n(t)) \log \frac{k_r(t)}{k^0_r} dt  \right]\\
 	&= \mathbb{E}_{k(\cdot)}\left[ \int_0^T C(n(t), k(t))dt \right].
\end{aligned}
\end{equation}
We can get the same expression by evaluating the KL divergence in small time intervals as in \cite{opper-2007-VariationalInferenceMarkov,nakamura-2022-OptimalSensingControl}.

It should be noted that the reaction count $\xi(t)$ contains more information than the number $n(t)$. 
For any $t$, the filtration $\mathcal{F}_t^n = \sigma(\{n(\tau)\}_{0 \leq \tau < t})$ generated by $n(\cdot)$ is the subset of the filtration $\mathcal{F}_t^\xi = \sigma(\{\xi(\tau)\}_{0\leq \tau < t})$, i.e., $\mathcal{F}_t^n \subset \mathcal{F}_t^\xi \subset \mathcal{F}_t$. 
This is because $n(t)$ can be calculated as a deterministic function of $\xi(t)$, specifically $n(t)=n(0)+\sum_r s_r \xi_r(t)$. In general, it is not possible to invert this function.
More precisely, if the stoichiometric matrix $S$ has a non-trivial right kernel, i.e., $\Ker S \neq \{0\}$, then $n(t)$ contains strictly less information than $\xi(t)$, meaning that $\mathcal{F}_t^n$ is a proper subset of $\mathcal{F}_t^\xi$.

In addition, by the data processing inequality, the KL divergence calculated for $n(\cdot)$ is smaller than the one calculated for $\xi(\cdot)$. Thus, the minimum control cost or minimum KL divergence problem for $\xi(\cdot)$ provides the upper bound for the minimum KL divergence problem for $n(\cdot)$.

\subsection{Solution of the optimal measure problem}\label{sec:proof_optimal_measure}
In this section, we present the proof of Eqs.~\eqref{eq:value_for_optimal_measure} and \eqref{eq:optimal_measure}, that is, the optimal probability measure problem has the same solution as the optimal control problem.

Assuming that $\mathbb{P}$ is absolutely continuous with respect to $\mathbb{P}_{k^0}$ and the KL divergence is finite, then we have
\begin{equation}
\begin{aligned}
	&\beta \mathbb{E}_{\mathbb{P}} \left[ \mathcal{U}(n(\cdot)) \right] - \mathcal{D}_{KL} \left[ \mathbb{P} \| \mathbb{P}_{k^0} \right]\\
	&= \mathbb{E}_{\mathbb{P}}\left[\beta \mathcal{U}(n(\cdot)) - \log \frac{d\mathbb{P}}{d\mathbb{P}_{k^0}}(\xi(\cdot))\right]\\
	&= \int \log\left(\exp(\beta \mathcal{U}(n(\cdot)))\frac{d\mathbb{P}_{k^0}}{d\mathbb{P}}(\xi(\cdot))\right) d\mathbb{P}(\xi(\cdot))\\
	&\leq \log\left(\int \exp(\beta \mathcal{U}(n(\cdot)))  \frac{d\mathbb{P}_{k^0}}{d\mathbb{P}}(\xi(\cdot)) d\mathbb{P}(\xi(\cdot))\right)\\
	&= \log\left(\int \exp(\beta \mathcal{U}(n(\cdot))) d\mathbb{P}_{k^0}(\xi(\cdot))\right)\\
	&=\log\mathbb{E}_{k^0}\left[ \exp(\beta\mathcal{U}(n(\cdot)))\right],
\end{aligned}
\end{equation}
where we used Jensen's inequality for the concave function $\log$. The equality holds if and only if $\exp(\beta\mathcal{U}(n(\cdot)))\frac{d\mathbb{P}_{k^0}}{d\mathbb{P}}(\xi(\cdot))$ is constant with probability one with respect to $\mathbb{P}$, which is given by the following $\mathbb{P}=\mathbb{P}^\dagger$:
\begin{equation}\label{eq:optimal_measure_re}
	d\mathbb{P}^\dagger(\xi(\cdot))=\frac{1}{\mathcal{Z}}\exp(\beta \mathcal{U}(n(\cdot)))d\mathbb{P}_{k^0}(\xi(\cdot)),
\end{equation}
where $\mathcal{Z}=\mathbb{E}_{k^0}\left[\exp(\beta\mathcal{U}(n(\cdot)))\right]$ is the normalizing constant.
Therefore, the optimum of the objective function is given by
\begin{equation}
\begin{aligned}
	\mathcal{V}(n_0, \beta) 
	&= \max_{\mathbb{P}} \left[\beta \mathbb{E}_{\mathbb{P}} \left[ \mathcal{U}(n(\cdot)) \right] - \mathcal{D}_{KL} \left[ \mathbb{P} \| \mathbb{P}_{k^0} \right]\right] \\
	&= \log\mathbb{E}_{k^0}\left[\exp(\beta\mathcal{U}(n(\cdot)))\right],
\end{aligned}
\end{equation}
which is equal to the optimum $V(n_0, \beta)$ of optimal control problem.

Furthermore, we can confirm that the optimally controlled process with $k^\dagger(t, n, \beta)$ has the probability law given in Eq.~\eqref{eq:optimal_measure_re}. Let us denote the set of times at which jumps occur as $\{\tau_i\}_{i=1, \ldots, I}$ where $\tau_0:=0\leq \tau_1 < \tau_2 < \cdots < \tau_{I-1} < \tau_{I} \leq T =: \tau_{I+1}$ and $I = \sum_{r\in R} \xi_r(T) \geq 0$ is the total number of jumps in the time interval $[0, T]$.
We further denote the reaction count and the state after $i$-th jump as $\xi^i := \xi(\tau_i)$ and $n^i := n(\tau_i)$, where we set $\xi^0 := \xi(0)$ and $n^0 := n(0)$ for convenience.
Since no two reactions occur simultaneously, the reaction occurs at time $\tau_i$ is denoted by $r_i\in R$, so that $\xi_{r_i}^{i} = \xi_{r_i}^{i-1}+1$ and $n^i = n^{i-1} + s_{r_i}$ hold.

Using Eq.~\eqref{eq:radon-nikodym}, the Radon–Nikodým derivative is given by
\begin{equation}\label{eq:radon-nikodym-optimal}
\begin{aligned}
	\log\frac{d\mathbb{P}_{k^\dagger(\cdot)}}{d\mathbb{P}_{k^0}}(\xi(\cdot))
	&= 
	-\int_0^T \sum_{r \in R} \left(k_r^\dagger(t, n(t), \beta) - k^0_r \right) h_r(n(t)) dt\\
	&+\int_0^T \sum_{r \in R} \log \frac{k_r^\dagger(t, n(t^-), \beta) }{k^0_r} d\xi_r(t).
\end{aligned}
\end{equation}

For the first term of Eq.~\eqref{eq:radon-nikodym-optimal}, we get
\begin{equation}
\begin{aligned}
	-\int_0^T \sum_{r \in R} \left(k_r^\dagger(t, n(t), \beta) - k^0_r \right) h_r(n(t)) dt
	&= -\int_0^T \sum_{r \in R} k^0_r\left(\exp(\overline{\nabla}_{s_r}V_t(n(t), \beta)) -  1\right) h_r(n(t)) dt\\
	&= \int_0^T \left(\frac{\partial}{\partial t}V_t(n(t), \beta) + \beta U_t(n(t))\right)dt\\
	&= \beta \int_0^T U_t(n(t))dt + \sum_{i=0}^I \int_{\tau_i}^{\tau_{i+1}} \frac{\partial}{\partial t}V_t(n^i, \beta)dt\\
	&= \beta \int_0^T U_t(n(t))dt + \sum_{i=0}^I \left(V_{\tau_{i+1}}(n^i, \beta) - V_{\tau_i}(n^i, \beta)\right),
\end{aligned}
\end{equation}
where we used the HJB Eq.~\eqref{eq:HJB_KL_appendix} in the second line.

For the second term of Eq.~\eqref{eq:radon-nikodym-optimal}, we get
\begin{equation}
\begin{aligned}
	\int_0^T \sum_{r \in R} \log \frac{k_r^\dagger(t, n(t^-), \beta) }{k^0_r} d\xi_r(t)
	&= \sum_{i=1}^I \log \frac{k_{r_i}^\dagger(t, n^{i-1}, \beta) }{k^0_{r_i}}\\
	&= \sum_{i=1}^I \overline{\nabla}_{s_{r_i}}V_{\tau_i}(n^{i-1}, \beta)\\
	&= \sum_{i=1}^I \left(V_{\tau_i}(n^{i-1}+s_{r_i},\beta) - V_{\tau_i}(n^{i-1}, \beta)\right)\\
	&= \sum_{i=1}^I \left(V_{\tau_i}(n^{i},\beta) - V_{\tau_i}(n^{i-1}, \beta)\right)
\end{aligned}
\end{equation}

Summing up both terms, the Radon–Nikodým derivative becomes
\begin{equation}
\begin{aligned}
	\log\frac{d\mathbb{P}_{k^\dagger(\cdot)}}{d\mathbb{P}_{k^0}}(\xi(\cdot))
	&=\beta \int_0^T U_t(n(t))dt + \sum_{i=0}^I \left(V_{\tau_{i+1}}(n^i,\beta) - V_{\tau_i}(n^i,\beta)\right) + \sum_{i=1}^I \left(V_{\tau_i}(n^{i},\beta) - V_{\tau_i}(n^{i-1}, \beta)\right)\\
	&=\beta \int_0^T U_t(n(t))dt + V_{\tau_{I+1}}(n^I, \beta) - V_{\tau_0}(n^0,\beta)\\
	&=\beta \int_0^T U_t(n(t))dt + \beta U_T(n(T)) - V_{0}(n^0,\beta)\\
	&= \beta \mathcal{U}(n(\cdot)) - V(n^0, \beta),
\end{aligned}
\end{equation}
where we used the terminal condition $V_{\tau_{I+1}}(n^I, \beta)=V_T(n(T), \beta)=\beta U_T(n(T))$.
Therefore, the probability law $\mathbb{P}_{k^\dagger(\cdot)}$ for the optimal control has the same expression
\begin{equation}
	d\mathbb{P}_{k^\dagger(\cdot)}(\xi(\cdot))=\frac{1}{\mathcal{Z}}\exp(\beta \mathcal{U}(n(\cdot)))d\mathbb{P}_{k^0}(\xi(\cdot)),
\end{equation}
as the optimal probability measure $\mathbb{P}^\dagger$ in Eq.~\eqref{eq:optimal_measure_re}.

\section{Infinite horizon average cost problems}\label{appendix:average_cost}
In this section, we consider the following optimal control problem:
\begin{equation}
	V_*(n_0, \beta) := \max_{k(\cdot)} \lim_{T\rightarrow \infty} \frac{1}{T}\mathbb{E}_{k(\cdot)} \left[ \int_0^T (\beta U(n(t)) - C(n(t), k(t)))dt ~\big| n(0) = n_0\right].
\end{equation}
We can derive the following optimality equation~\cite{xianpingguo-2003-DriftMonotonicityConditions}: for all $n\in \mathcal{N}$,
\begin{equation}
	\chi(\beta) = \max_k \left[\beta U(n) - C(n, k) + \mathcal{L}_k v(n, \beta)\right],
\end{equation}
where $\chi(\beta)\in\mathbb{R}$ and a scalar function $v(n, \beta)$ on $\mathcal{N}\times\mathbb{R}$. 
If the instantaneous control cost function $C(n, k)$ is assumed to be continuous, strictly convex, and 1-coercive, we have
\begin{equation}
	\chi(\beta) = \beta U(n) + C^*(n, h(n) \overline{\nabla}_s v(n, \beta)).
\end{equation}
The optimum $V_*(n_0, \beta)$ is given by $V_*(n_0, \beta)=\chi(\beta)$ independently of the initial condition $n_0$. The optimal control $k^\dagger(n,\beta)$ is given by
\begin{equation}
	k^\dagger(n,\beta) = \partial_\phi C^*(n, h(n)\overline{\nabla}_s v(n, \beta)).
\end{equation}

When we further assume the KL control cost, the optimality equation becomes
\begin{equation}
	\chi(\beta) = \beta U(n) + \sum_{r\in R} k_r^0 h_r(n) \left(\exp(\overline{\nabla}_{s_r} v(n, \beta)) - 1\right).
\end{equation}
Applying Cole-Hopf transformation $z(n, \beta):=\exp v(n, \beta)$, we obtain
\begin{equation}
	\chi(\beta)z(n, \beta) = \beta U(n)z(n, \beta) + \mathcal{L}_{k^0} z(n, \beta).
\end{equation}
Thus, it is reduced to the eigenvalue problem, which is evident in the following matrix-vector representation:
\begin{equation}
	\chi(\beta)\bm{z} = (\beta \diag \bm{u} + \Omega^0)\bm{z},
\end{equation}
where $(\bm{z})_n=z(n,\beta)$, $(\bm{u})_n=U(n)$, and $(\Omega^0)_{n,n^\prime}=\sum_{r\in R} k^0_r h_r(n) \delta_{n+s_r, n^\prime}$.
Here $\chi(\beta)$ is the maximum eigenvalue and $\bm{z}$ is the corresponding eigenvector.
The eigenvectors $\bm{z}$ can have arbitrary scale, which is consistent with the fact that the constant shift of the function $v(n,\beta)$ does not affect the optimal control $k^\dagger_r(n, \beta)= k^0_r \exp(\overline{\nabla}_{s_r} v(n, \beta))$.
The eigenvalue problem can be solved efficiently if the state space $\mathcal{N}$ is bounded and finite.

Finally, we briefly mention another infinite horizon problem with the following discounted objective function:
\begin{equation}
	V_\alpha(n_0, \beta) := \max_{k(\cdot)} \mathbb{E}_{k(\cdot)} \left[ \int_0^\infty e^{-\alpha t}(\beta  U(n(t)) - C(n(t), k(t))) dt ~\big| n(0) = n_0\right],
\end{equation}
where $\alpha > 0$ is the discount factor.
The value function $V_\alpha(n, \beta)$ satisfies
\begin{equation}
	\alpha V_\alpha(n,\beta) = \beta U(n) + C^*(n, h(n) \overline{\nabla}_s V_\alpha(n, \beta)).
\end{equation}
When we assume the KL control cost, it becomes
\begin{equation}
	\alpha V_\alpha(n,\beta) = \beta U(n) + \sum_{r\in R} k_r^0 h_r(n) \left(\exp(\overline{\nabla}_{s_r} V_\alpha(n, \beta)) - 1\right).
\end{equation}
Applying the transformation $Z_\alpha(n, \beta):=\exp(V_\alpha(n, \beta))$, we obtain
\begin{equation}
	\alpha Z_\alpha(n,\beta)\log(Z_\alpha(n,\beta)) = \beta U(n) + \mathcal{L}_{k^0}Z_\alpha(n, \beta),
\end{equation}
which is not a linear equation with respect to $Z_\alpha$.
Therefore, the discounted cost problem cannot be solved efficiently, unlike the first exit problem or the average cost problem. The previous study by Todorov reported the analogous conclusion about solvability for diffusion processes and discrete-time Markov chains~\cite{todorov-2009-EfficientComputationOptimal}.

\section{\ADD{Previous studies}}\label{appendix:previous_studies}
\ADD{Here, we compare our results with existing literature.
Although the diffusion process is the most extensively studied continuous-time Markov process in optimal control problems, 
the problems for Markov jump processes have also been formulated by \cite{boel-1977-OptimalControlJump,pliska-1975-ControlledJumpProcesses,rishel-1975-MinimumPrincipleControlled,davis-1977-OptimalControlJump}, where the optimality conditions and existence and uniqueness of the solution were established. General results and historical remarks can be found in \cite{fleming-2006-ControlledMarkovProcesses,cohen-2015-StochasticCalculusApplications}.
}

\ADD{
Thereafter, many studies worked on the optimal control problems in complex settings like stochastic reaction networks in the context of chemical reactions \cite{guo-2009-OPTIMALCONTROLSTOCHASTIC,zhang-2015-OptimalControlMarkov,briat-2022-ContinuousTimeSampledDataOptimal,briat-2022-ContinuousSampledDataControl,briat-2021-OptimalH_inftyControl}, jump diffusions \cite{theodorou-2012-StochasticOptimalControl,okumura-2017-IterativePathIntegral}, ecology \cite{fischer-2015-ValueMonitoringControl}, epidemiology \cite{lefevre-1981-OptimalControlBirth,cai-1994-StochasticControlEpidemic,lorch-2018-StochasticOptimalControl} and economics \cite{jaimungal-2024-MinimalKullbackLeibler}.
However, there were no widely applicable efficient methods to solve the problems in complex settings like stochastic reaction networks.
}

\ADD{
Note that while the stochastic reaction network is a special class of Markov jump processes, stochastic reaction networks have several unique features that call attention to their optimal control problems.
First, it has special algebraic constraints in the jump size imposed by stoichiometry. If the stoichiometric matrix is not full rank, the reachability among the states is limited by the stoichiometric subspace \cite{feinberg-2019-FoundationsChemicalReaction,anderson-2011-ContinuousTimeMarkov}.
Second, in controlling stochastic reaction networks, the propensity function, which can have a nonlinear dependence on the state, governs how the control input affects the process. If all the reactions are unimolecular with mass action kinetics, the propensity function is linear. In this case, the optimal control problems with quadratic control cost can be solved analogously to linear quadratic problems \cite{briat-2022-ContinuousTimeSampledDataOptimal,briat-2022-ContinuousSampledDataControl,briat-2021-OptimalH_inftyControl}, which cannot be applied to reaction networks with nonlinear propensity functions.
Third, the size of the state space can be infinite and grow exponentially as the number of types increases. In the context of epidemics, for example, the number of types can be very large in order to capture the complex dynamics arising from heterogeneity in the population \cite{brauer-2012-MathematicalModelsPopulationa}.
Prior techniques for solving optimal control problems are limited in applicability to complex situations or efficiency in large-scale problems.
}

\ADD{
Path integral control and KL control concepts were found to be effective in solving optimal control problems only for diffusion processes and discrete-time Markov chains \cite{kappen-2005-PathIntegralsSymmetry,todorov-2009-EfficientComputationOptimal} in physics and computational neuroscience. As pointed out by \cite{theodorou-2012-RelativeEntropyFree}, a mathematically nice property of KL divergence on path probability measures is fundamental in path integral control for diffusion processes and KL control for discrete-time Markov chains.
The same discussion applies to continuous-time Markov jump processes, and several recent studies \cite{gao-2023-OptimalControlFormulation,gueant-2020-OptimalControlFinite,jaimungal-2024-MinimalKullbackLeibler} have presented similar theories. It is noteworthy that the basically same mathematical structure has already been introduced by Fleming \cite{fleming-1977-ExitProbabilitiesOptimal} and Holland \cite{holland-1977-NewEnergyCharacterization} in 1977 and its application in Markov jump process can be found in Fleming's 1982 paper \cite{fleming-1982-LogarithmicTransformationsStochastic, fleming-2006-ControlledMarkovProcesses}.
}

\section{Maximum speed problem for a random walker in a fixed duration}
\label{appendix:random_walk_finite_horizon}
In this section, we present another analytically solvable optimal control problem with a random walker on a line.
We consider the following problem
\begin{equation}
	\underset{k(\cdot)}{\text{maximize}} \quad  \beta \mathbb{E}_{k(\cdot)}\left[x(T)\right] - \mathcal{D}_{KL} \left[ \mathbb{P}_{k(\cdot)} \| \mathbb{P}_{k^0} \right],
\end{equation}
where $x(t) = \hat{x}(n) \in X$ represents the position of the particle at time $t$.
For simplicity, we assume the uncontrolled reaction rate coefficients or the uncontrolled transition rates are the same $k^0_r = \kappa > 0$ for all $r\in R$. Setting $U_T(n)= \hat{x}(n)$ and $U_t(n)\equiv 0$ and using the probabilistic representation, Eq~\eqref{eq:probabilistic_rep_value_reaction}, the value function $V_t(n, \beta) =: V_t(\hat{x}(n), \beta)$ is given by
\begin{equation}
	V_t(x,\beta) = \log\mathbb{E}_{k^0}\left[\exp(\beta x(t))|x(0)=x\right].
\end{equation}
From the analytical expression of the probability distribution~\cite{feller-1991-IntroductionProbabilityTheory_vol1}, we obtain
\begin{equation}
	V_t(x, \beta) = \beta x + 2\kappa(T-t)(\cosh(\beta)-1).
\end{equation}
Since the value function is an affine function in $x$, the optimal reaction rate coefficients are independent of the position $x$ and time $t$,
\begin{equation}
\begin{aligned}
	k^\dagger_{(x,x+1)}(t, n, \beta) &= \kappa e^\beta,\\
	k^\dagger_{(x,x-1)}(t, n, \beta) &= \kappa e^{-\beta}.
\end{aligned}
\end{equation}
Under this optimal control, the expected position at the terminal time can be calculated using Eq.~\eqref{eq:optimal_utility_appendix}:
\begin{equation}
	\mathcal{U}_t^\dagger(x, \beta) := \mathbb{E}_{k^\dagger}[x(t) |x(0)=x] =
	x + 2\kappa  (T-t) \sinh(\beta).
\end{equation}
The first term $x$ is the expected utility for uncontrolled processes $\mathcal{U}_t^\dagger(x, 0)=x$.
The second term $2\kappa  (T-t) \sinh(\beta)$ is gained by consuming the control cost
\begin{equation}
	\mathcal{C}^\dagger_t(x, \beta) = 2\kappa (T-t) (\beta \sinh(\beta)-\cosh(\beta)+1),
\end{equation}
which can also be represented by the conjugate value function
\begin{equation}\label{eq:conjugate_value_random_walk_finite_time}
	\mathcal{E}_t(x, u) = 2\kappa (T-t) \left(\hat{u}\sinh^{-1}(\hat{u})-\sqrt{1+\hat{u}^2} + 1\right),
\end{equation}
where $\hat{u} = \left(u - x\right) / (2\kappa (T-t))$.

\section{Additional results on controlling Moran processes}
\label{appendix:moran_control}

\begin{figure*}[htp]
\centering
\includegraphics[width=15cm]{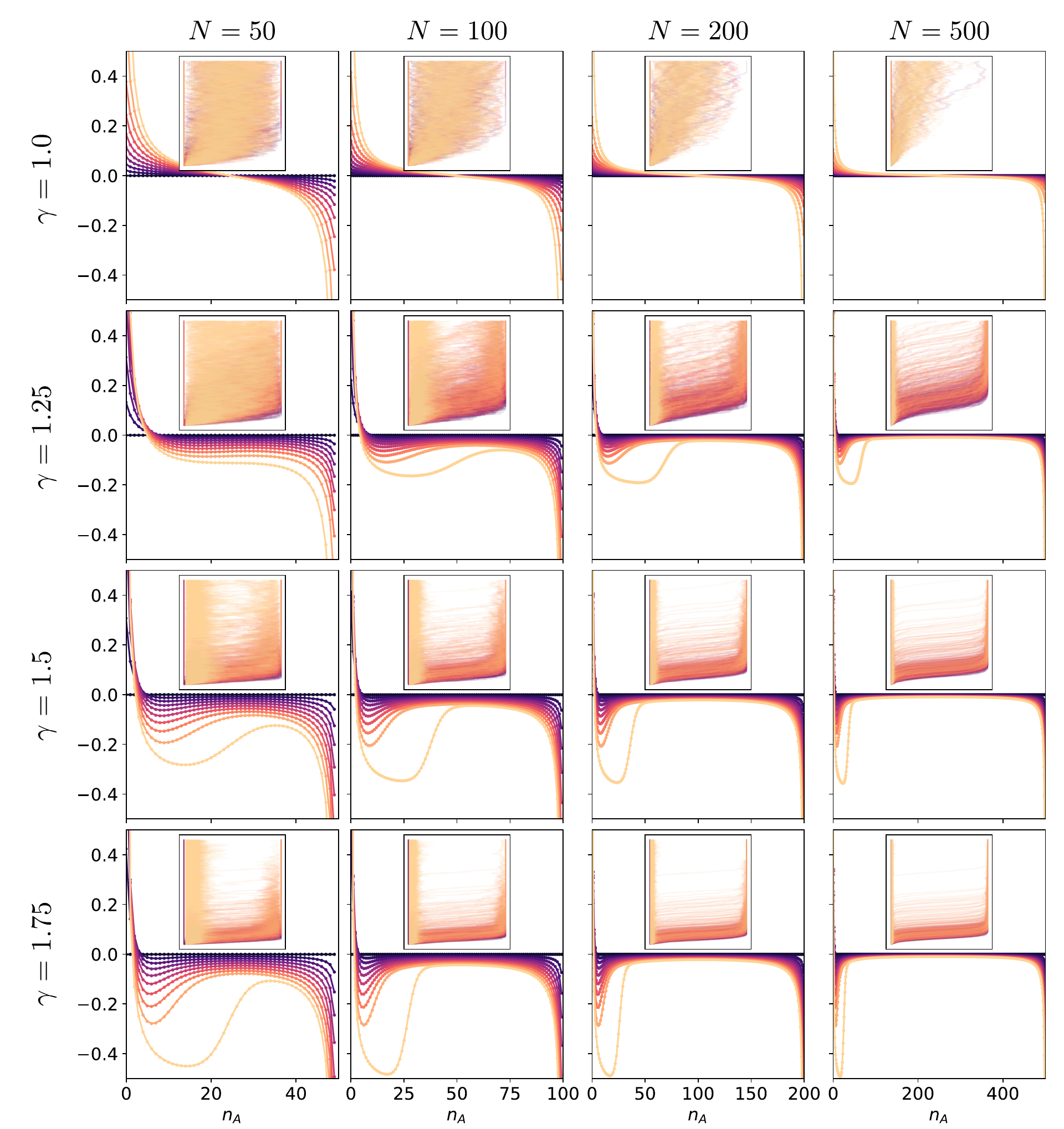}
\caption{The results of the maximum exit time problem for Moran processes with different $N$ and $\gamma$. Each panel shows the dependence of the derivative $\overline{\nabla}_{s_1}V(n,\beta)$ of the value function on the initial state $n_A$ for different $\beta$. The values of $\beta$ are sampled at equal intervals between $0$ to $\beta_c$ and color-coded. The optimally controlled stochastic trajectories from time $0$ to $T=100$ with initial condition $n_A(0)=1$ for different $\beta$ are also shown in the inset.}
\label{fig:Moran_prameter_region}
\end{figure*}

In the main text, we calculated the results of the maximum exit time problem for Moran processes with population size $N=100$. Here, we show additional results with smaller and larger population sizes $N=50, 100, 200$, and $500$ in Fig.~\ref{fig:Moran_prameter_region}.

When $N$ or $\gamma$ is small, the derivative of the value function does not have the plateau region (OFF mode region), and the controlled trajectories can maintain an intermediate size of $n_A$.
On the other hand, when $N$ and $\gamma$ are large, the OFF mode region emerges, and the controlled trajectories show bimodality.

Focusing on the size of the ON mode region around $n_A=0$, its relative size decreases as $N$ increases. However, the absolute size of the ON region remains unchanged.
Therefore, we expect that the optimal control law becomes singular in the large population limit $N\rightarrow \infty$ where the ON mode region appears only at the boundary $n_A=0$.

\bibliography{ocsrn}

\end{document}